\documentclass[11pt,preprint]{aastex}
\def\ltsim{\raise 2pt \hbox {$<$} \kern-1.1em \lower 4pt \hbox {$\sim$}}

\slugcomment{ApJ in press - 10 Jan 2005 issue}
\shortauthors{Giovannini G., Taylor G.B., Feretti L., et al.}
\shorttitle{A Complete VLBI Sample of Radio Sources} 

\begin{document}

\title{The Bologna Complete Sample of Nearby Radio Sources}

\author{G. Giovannini\altaffilmark{1,2}, G.B. Taylor\altaffilmark{3},
L. Feretti\altaffilmark{2}, W.D. Cotton\altaffilmark{4}, 
L. Lara\altaffilmark{5,6}, and T. Venturi\altaffilmark{2}}
 
\altaffiltext{1}{Dipartimento di Astronomia, Universita' di Bologna,via 
Ranzani 1, 40127 Bologna, Italy} 
\altaffiltext{2}{Istituto di Radioastronomia del CNR, via Gobetti 101, 40129
Bologna, Italy}
\altaffiltext{3}{National Radio Astronomy Observatory, PO Box O, Socorro NM
87801, USA}
\altaffiltext{4}{National Radio Astronomy Observatory, 520 Edgemont Rd,
Charlottesville \\
VA 22903-2475, USA}
\altaffiltext{5}{Dpto Fisica Teorica y del Cosmos, Universidad de Granada,
avda Fuentenueva s/n, 18071 Granada, Spain}
\altaffiltext{6}{Instituto de Astrofisica de Andalucia, CSIC. Apdo. 3004,
18080 Granada, Spain}
\email{ggiovann@ira.cnr.it,gtaylor@nrao.edu,lferetti@ira.cnr.it,
bcotton@nrao.edu,lucas@ugr.es,tventuri@ira.cnr.it}

\begin{abstract}
We present a new, complete, sample of 95 radio sources selected from the
B2 and 3CR catalogues, with z $<$ 0.1.  Since no selection effect on
the core radio power, jet velocity, or source orientation is present,
this sample is well suited for statistical studies.  
In this first
paper we present the observational status of all sources on the 
parsec (mas) and kiloparsec (arcsec) scale; 
we give new parsec-scale data for 28 sources
and discuss their parsec-scale properties. 
Combining these data with that in the literature, information on the 
parsec-scale morphology is available for a total of 
53 radio sources with different radio power and kpc-scale morphology.
We investigate their properties. We find a dramatically higher fraction of
two-sided sources in comparison to previous flux limited VLBI surveys.

\end{abstract}


.
 
\vfill
\eject

\section{Introduction}

The study of the parsec scale properties of radio galaxies is crucial
to obtain information on the nature of their central engine, and
provides the basis of the current {\it unified theories} 
\citep[see e.g.][]{ur95}, 
which suggest that the appearance of
active galactic nuclei strongly depends on orientation. In the {\it
high-luminosity unified scheme}, quasars and powerful FR II radio
galaxies should be the same class of objects seen at different viewing
angles.  Similarly, the {\it low-luminosity unified scheme} assumes
BL-Lacs to be the beamed population of radio galaxies of
low-intermediate luminosity (FR I).

To gain new insights in the study of radio galaxies on pc scales,
it is important to select source samples from low
frequency catalogues, where the source properties are dominated by the
unbeamed extended emission and are not affected by observational
biases related to orientation effects.  To this aim, we undertook a
project to observe a complete sample of radio galaxies
selected from the B2 and 3CR catalogs \citep{g90,g01}.  
However, because of obvious observational
limitations at the time of the sample selection, we chose radio
sources with a relatively high core flux density ($>$ 100 mJy at
6 cm).  This clearly favors objects with a beamed core, i.e., oriented
at a small angle with respect to the line of sight. We will refer to
this sample as the {\it Bologna Strong Core Sample} (BSCS).

To overcome this bias, we recently selected a {\bf second sample} with
{\bf all} the B2 and 3CR radio sources present in the same sky region
of the first sample with z $<$ 0.1, regardless of the core flux
density: the {\it Bologna Complete Sample} (BCS).  
This sample
consists of 95 sources and includes all the sources of the BSCS except
3C109, 3C 303, and 3C 346 because of the redshift limit. At present 53
of the 95 sources have been studied with VLBI observations.  Here we
present this new sample, report pc-scale data for the 53 sources with
VLBI observations, and discuss in detail the structure and properties
for the 28 sources with new VLBI data.  In the near future we plan to
observe the 42 sources not yet observed with parsec resolution using the
phase-referencing technique since the nuclear flux density in these
sources is too low to be detected using the standard technique.

For consistency with previous studies we adopt a Hubble constant 
H$_0$ = 50 km sec$^{-1}$ Mpc$^{-1}$ and q$_0$ = 0.5 ($\Omega_0$ = 1). 
Because of the low redshift
limit the adopted value of the cosmological constant 
$\Omega$ is not relevant.

\section {The Sample}

In Table 1 we give the complete list of radio sources and summarize
the most relevant information. The sample, consisting of 95 radio
galaxies and BL-Lac type objects, was obtained from the B2 and 3CR
catalogues. We used the sources above a homogeneous flux density limit
of 0.25 Jy at 408 MHz for the B2 sources and above 10 Jy at 178 MHz for
the 3CR sources \citep{fe84}, and applied the following
criteria:

\begin{itemize}
\item Declination $>$ 10$^\circ$
\item Galactic latitude $|$ b $|$ $>$ 15$^\circ$
\item Redshift z $<$ 0.1
\end{itemize}

For all sources  high quality images
at arcsecond resolution obtained with the Very Large Array (VLA)
of the NRAO\footnote {The National Radio Astronomy 
Observatory is operated
by Associated Universities, Inc., under cooperative agreement with the
National Science Foundation.} are available in the literature, allowing
a detailed study of the large scale structure.
According to the kpc-scale morphology the sample contains: 65 FR
I radio galaxies; 15 FR II sources; 13 compact sources, 
and 1 FRI/II source. One more source has been tentatively classified
as a spiral galaxy according to the optical and large scale
radio strucure. If confirmed it will not belong to the present sample.
Moreover, VLA observations at 5 GHz provide a measure
of the nuclear emission at arcsecond resolution which we refer to as 
the arcsecond
core \citep[see][]{g88}, while we use the term VLBI core to refer to the
unresolved compact
component visible in the VLBI images at milliarcsecond resolution.

\section{Observations}

We obtained VLBA observing time at 5 GHz, in order to produce high resolution
images for 28 sources of the BCS. For these observations we selected
all sources with an estimated nuclear flux density high enough ($\sim$
10 mJy or more) to allow standard VLBI observations (without 
phase-referencing). Sources with a lower core flux density will be
observed in a second session.  Observations were made with the full
VLBA of the NRAO
on April 4, 1997 (12 hours of observing time) and on January 22,
2000 (24 hours of observing time). Each source was observed for about
1 hour with short scans at different hour angles to ensure good
uv-coverage. The observations were correlated in Socorro,
NM. Postcorrelation processing used the NRAO AIPS package. All data
were globally fringe fitted and self-calibrated. 
Final images were obtained using the AIPS and DIFMAP packages.  
The parameters of the observed sources are presented in Table 2. The VLBI core
properties have been estimated by fitting a Gaussian to the VLBI
image. The total VLBI flux is the correlated flux present on the short
baselines. The noise level was estimated from the final images. 
The differences in measured noise properties are partially due to the two
different observing sessions. 
Frequent observations of OQ208 were used to calibrate polarization terms.  
Images in
polarization mode were obtained but no sources were detected. Limits on
the polarized flux percentage are given in Table 2 as the ratio
between 3 $\sigma$ level in the polarized map and the peak flux in the
total intensity image.

\section{Notes on Individual Sources}

In the following, we provide short descriptions of all the sources
presented in this paper, together with some information on the large
scale structure. We present contour images for the resolved sources.

{\bf 0106+13 -- 3C33} (Figure 1): On the kpc-scale this narrow line FR II 
radio
source shows a symmetric structure with two extended lobes and bright
hot spots \citep{lea91}. The VLBI image shows
nuclear emission with a flux density of $\sim$ 10.6 mJy and two
symmetric jets aligned with the kpc-scale structure. The southern jet
is stronger and the brightness ratio to the counter-jet is $\sim$2 at
$\sim$5 mas from the core and $\sim$1 at 10 mas, suggesting a jet velocity 
decrease.

{\bf 0326+39} (Figure 2): This is a symmetric FR I radio galaxy 
extended in the
E--W direction, with the main jet on the western side \citep{br91}.
In our VLBI image a dominant nuclear emission is
seen with a short, faint jet on the same side as the main kpc-scale
jet. A faint extension at ~~$\ltsim$ ~ 2$\sigma$ level is present on the
opposite side and could be a marginal detection of the counter-jet.

{\bf 0844+31 -- 4C31.32} (Figure 3): This is a symmetric narrow line 
FR II radio
galaxy.  At pc resolution a bright core and two-sided jets are
visible.  The symmetric pc-scale structure and the absence of a
visible broad line region
suggest that this source is at a large angle with respect
to the line-of-sight. However, the core radio power is comparable to
the total radio power, suggesting a moderate angle to the
line-of-sight.  The image from the VLA Faint
Images of the Radio Sky at Twenty-cm Survey \citep[FIRST,][]{first}
shows two extended FR I lobes at a larger distance from the
core beyond the FR II type hot spot, indicating a
prior phase of radio activity.  This source could be classified 
as a re-started source, i.e., a source where the extended FR I structure is
related to previous activity whereas the inner FR II structure
originates from more recent core activity.  This scenario explains
the apparent contradiction of a core dominated source with a symmetric pc scale
jet, since it is not
expected for the present core radio power to correlate with the total
source radio power: the present core dominance and the powerful parsec-
scale structure could be related to a recent commencement of a third
phase of increased activity. Multi-epoch observations could detect a possible
growth of the inner jets allowing a more detailed study of this
interesting source.

{\bf 1003+35 -- 3C236} (Figure 4): This giant radio galaxy, with its projected
linear size of more than 6 h$_{50}^{-1}$ Mpc, is the largest known
radio galaxy. It has been studied on a vast range of angular scales
and frequencies and it shows a complex central structure at arcsecond
resolution
\citep{fo82,ba85}. 
Using recent VLBI observations, \citet{sc01,tay01}
discussed the structure of the
steep spectrum core in detail and showed that this source has relativistic
jets oriented in the same direction as the large scale structure with
some oscillations (an ``S'' shaped structure is visible in the VLBA
images by \citet{tay01}).  Its peculiar radio structure has been
interpreted as evidence of re-started activity \citep{od01}.

The parsec scale structure is clearly asymmetric. \citet{sc01}
measured a jet/cj ratio of $\sim$ 50 at 1.6 GHz. In our 5.0 GHz
images we find a ratio $>$ 45 in the inner jet structure
(within 10 mas from the core). This strong asymmetry implies a source
orientation $\theta$ $<$ 50$^\circ$ with respect to the line of
sight. Taking into account the large size of this source we will
assume $\theta$ $\sim$ 50$^\circ$ which implies a highly relativistic
jet velocity ($\gamma$ $>$ 5). \citet{sc01} considered
optical data \citep{dk00} and suggested an average radio axis
for this source at $\sim$ 60$^\circ$ with respect to the line of
sight.  This value is not too far from the value derived from the jet
counter-jet
ratio taking into account the jet oscillations visible in high
resolution images \citep{tay01}.  In our VLBA images, despite
of the limited uv coverage, the core
is readily visible, as well as the one-sided NW jet, the compact NW blob, and
the extended emission at $\sim$ 100 mas from the core in the SE
direction.

{\bf 1037+30} (Figure not included): The VLA image shows a compact double 
structure
2$^{\prime\prime}$ in size, but the position of the nucleus is
uncertain.  Our VLBI observation were pointed at the peak emission in
the VLA image \citep{fan87}. The source was not detected ($<$ 2
mJy). This could be due to an incorrect core position and/or the
presence of extended sub-arcsecond structure.

{\bf 1040+31} (Figure 5): This source has a peculiar arcsecond scale radio
morphology: it shows an extended halo with a jet connecting the core
with a faint hot spot \citep{fan87}.  In our VLBI image a pc-scale 
jet is oriented at the same PA as the kpc-scale jet. A symmetric
faint cj is visible ($\sim$ 2$\sigma$ level).  From the core dominance
the source should be oriented in the range 40$^\circ$ -- 50$^\circ$
with a high velocity jet.  The symmetric pc-scale structure, if
confirmed, implies an orientation of $\sim$ 50$^\circ$ - 70$^\circ$.

{\bf 1102+30} (Figure 6): On the kpc-scale this source is a symmetric 
FR I radio
galaxy. From the FIRST image there is a hint that the main jet is on
the W side (see Fig. 6--left).  At pc-scale resolution we find a core and a
one-sided jet on the W side, confirming that the W jet is the
approaching one.  On the E side a low level emission may be
present.

{\bf 1116+28} (Figure 7): At arcsecond resolution this source shows a Narrow
Angle Tail (NAT)
structure with two symmetric jets.  Our VLBA image is
slightly resolved with a two-sided jet structure oriented in the
same directions as the kpc-scale jets.

{\bf 1204+34} (Figure 8): This FR II radio galaxy shows on the kpc-scale 
an 'X' shaped structure
with a central core and possibly a one-sided jet to the NW
\citep{fan87}. Our VLBI image shows the core and a short
one-sided jet at about the same PA as the kpc-scale, jet-like feature.

{\bf 1316+29 -- 4C29.47} (Figure 9): The pc-scale image shows a symmetric
extended structure oriented East--West.  The VLA image shows
extended emission aligned with the pc-scale structure, however the
two-sided jet emission at arcsecond resolution is in the N-S direction
with a clear 'S' shaped structure and a 90$^\circ$ jet curvature
\citep{fan87}.

{\bf 1321+31 -- NGC5127} (Figure not included): On the kpc-scale a core and 
a two-sided
symmetric jet is present. In our VLBA image
only a faint unresolved nuclear component (3.7 mJy) is present.
\citet{xu00} found at 1.67 GHz a
two-sided jet aligned with the kpc-scale emission.
On our shortest baseline a total flux density of 11 mJy is present, suggesting
the presence of a larger scale structure not visible in our images. We note
that self-calibration is not reliable for this source
because of the too low flux density.

{\bf 1346+26 -- 4C26.42} (Figure not included): This source is 
identified with the central cD
in the cooling flow cluster A1795. In the VLA images it is a slightly
extended (20'') with a FR I 'Z' shaped structure.  Its
interaction with the X-ray emitting gas visible in Chandra images
has been recently discussed in detail \citep{fa01}.

We marginally detect this source in our VLBI data.  The structure is
poorly defined, but possibly extended with a total flux density of
$\sim$8 mJy.  The large difference between the arcsecond core flux
density (53 mJy) and the flux density measured on the mas scale suggests
the presence of extended emission with a complex morphology. Deeper
observations and better (u,v) coverage will be necessary to
properly image this source.

{\bf 1350+31 -- 3C293} (Figure 10): This source has been studied in detail at
arcsecond and sub-arcsecond resolution \citep[][and
references therein]{be04}.  Here we present a pc-scale image of
the core region where a two-sided jet structure is aligned with the
sub-arcsecond scale structure. A large amount of correlated
flux is present only on the shortest baseline (Y - VLBA-PT) but it is
missing in our image because of the poor (u,v) coverage. The symmetric
structure suggests that the present active jets are nearly in the
plane of the sky. The connection with the kpc-scale structure is
not clear.  It could be a case of a restarted activity as argued for 
0844+31,
1003+36, and 1842+45. In this case the recent emission appears to be at a
different PA with respect to the older emission.

{\bf 1414+11 -- 3C296} (Figure 11): This symmetric FR I radio galaxy shows
two-sided emission in our VLBI image as well as in the VLA
images \citep{lea91}. From polarization observations
\citep{ga96} the Southern lobe appears to be the more
distant one (due to the Laing-Garrington effect).  See also
the Leahy et al. web page:
http://www.jb.man.ac.uk/atlas).  This is in agreement with the
evidence that the brighter, kpc-scale jet points to the north, and with
the brightness asymmetry
in our VLBI image even if the low jet counter-jet ratio suggests 
an orientation close to the
plane of the sky.  A second epoch observation did not show any visible
proper motion in this smooth radio jet.

{\bf 1422+26} (Figure 12): This extended FR I galaxy shows at arcsecond
resolution two symmetric jets oriented E-W, with the main jet
on the western side.  This source is very faint on the pc-scale
(peak flux = 1.7 mJy/beam) and shows a possible one-sided jet at the
same PA as the main kpc jet. If this structure is real the
source is oriented at $\theta$ = 45$^\circ$ -- 50$^\circ$ and $\beta$
$>$ 0.95. More observations are necessary to confirm this result.

{\bf 1448+63 -- 3C305} (Figure not included): This is a small FR I 
radio galaxy showing a 'plumed'
double structure, with two faint hot-spots and symmetric jets \citep{he82}.
The optical galaxy is peculiar with continuum emission
on the HST scale perpendicular to the radio jet \citep{ja95}.
We marginally detected this FR I source with a possible extension
perpendicular to the kpc-scale jet, i.e., aligned with the HST
emission. The low correlated flux requires a deeper image to properly
study this source.

{\bf 1502+26 -- 3C310} (Figure 13): This powerful FR I radio source shows two
symmetric extended lobes oriented in the N-S direction. No kpc-scale
jets are visible in the available images. It was observed with a global
array at 18 cm \citep{gi02} and a peculiar pc-scale
structure was found: two compact components almost perpendicular to
the extended structure and an extended emission to the north, but
misaligned by about 20$^\circ$ with respect to the extended lobes. 
Our 6 cm image is inconsistent 
with this result since it shows a core and a one-sided
jet to the South, aligned
with the kpc-scale extended emission.  According to
observational data, if the pc-scale jets are highly relativistic
($\gamma$ $>$ 3) the source orientation is $\theta$ = 50$^\circ$ --
70$^\circ$.

{\bf 1521+28} (Figure 14): This extended FR I source shows a well 
defined one-sided
jet emission in VLA images \citep{ca95}. The same
structure is visible in our VLBA image where a short one-sided jet
is found in alignment with the kpc-scale jet.

{\bf 1529+24 -- 3C321} (Figure not included): This Narrow Line FR II 
radio galaxy shows two
symmetric lobes with hot spots and a faint jet in the NW
direction. Only a point-like core is visible in our VLBA map.
From the core dominance, we find that high velocity
jets are in agreement with observational data only if $\theta$ is
45$^\circ$ -- 70$^\circ$. This relatively large orientation angle is
consistent with the lack of a visible broad line region.

{\bf 1553+24} (Figure 15): At arcsecond resolution this source shows a well
defined high brightness main jet and a faint counter-jet. An optical
jet has been found coincident with the main kpc-scale jet \citep{ca00}.
Our mas scale image shows a one-sided jet structure 
aligned with the optical and large-scale jet.  We
derive from our data an inclination angle $\theta$ = 30$^\circ$
-- 40$^\circ$ in agreement with the presence of an optical jet.

{\bf 1557+26 -- IC4587} (Figure 16): In the FIRST image there is an extended
emission with a total flux density of 32.4 mJy. The host galaxy is
identified as a low redshift (0.0442) elliptical galaxy \citep{pa86}.  
A nearby point-like source with a flux density of 120 mJy
is present at about 30'' east of the core.  
The two sources are probably
unrelated therefore the radio galaxy IC4587 is below the B2 flux
density limit (because the large beam in the B2 catalogue blended the
two sources). The VLBA component associated with IC4587 is slightly
extended in the same direction of the arcsecond structure visible in
the FIRST image. The data presented here are for IC4587
only.

{\bf 1613+27} (Figure not included): A typical FR I radio galaxy. 
In our VLBA image we detect only 
unresolved nuclear emission.

{\bf 1621+38 -- NGC6137} (Figure 17): This head-tail radio galaxy 
has bright, short
symmetric jets oriented N-S on the kpc-scale, and an extended low brightness 
tail to
the east \citep{fan87}. In our VLBI image a bright core and a
one-sided jet is present.  The pc-scale jet is slightly
misaligned, but oriented on the same side with respect to the core as
the brighter kpc-scale jet ($\sim 160^\circ$ VLBI, $\sim 190^\circ$
VLA).

{\bf 1658+30 -- 4C30.31} (Figure 18): On the kpc-scale this source has a 
symmetric
FR I morphology with two extended lobes.  The main jet points to the
SW and is easily visible while the counter-jet is embedded in the
extended low brightness emission of the lobe \citep{fan87}. On
the parsec scale we detect a one-sided jet structure oriented
in the same direction as the main arcsecond-scale jet.

{\bf 1827+32A} (Figure 19): This FR I radio galaxy has an 
S-shaped structure.  In
the VLA high resolution image \citep{fan87}, the eastern jet is
the main one and a faint counter-jet is also apparent.  In our VLBI
image we detect a core and a one-sided jet with the same
orientation as the arcsecond-scale jet.

{\bf 1842+45 -- 3C388} (Figure 20): This source has been classified as 
an FR II
source identified with a bright cD galaxy, but its radio morphology is
quite peculiar: the eastern lobe is relaxed, while the western lobe
shows extended structure beyond the hot spot (as in 0844+31),
suggesting restarted activity (see Leahy et al.  home page:
http://www.jb.man.ac.uk/atlas).  On the basis of the lack of
depolarization asymmetry in the two lobes, \citet{ro94}
suggest that this source is oriented close to the plane of the sky.

Our VLBI image shows a core and one-sided jet emission with the same
orientation as the kpc-scale jet.  Observational data
suggest the presence of a relativistic jet oriented in the range
45$^\circ$ to 65$^\circ$.  We note that an angle of $\sim$ 60$^\circ$
is in agreement with \citet{ro94} interpretation of the
polarization data.

{\bf 1855+37} (Figure not included): This low power compact source has a 
distorted double structure
on the kiloparsec-scale.
We did not detect this source in our VLBI observations pointed in between the
two arcsecond lobes suggesting the identification as 1855+37 as a 
small symmetric source with a faint core.

{\bf 2229+39 -- 3C449} (Figure 21): We detected only a faint, unresolved
core, in agreement with the orientation near to the plane of the sky of
this well known FR I source \citep[see][]{fe99}.

\section {Source Morphology}

Among sources with VLBI data we have the following kpc-scale structures:

\begin{itemize}

\item
31 FR I radio galaxies

\item
10 FR II radio galaxies

\item
8 Compact sources with a flat spectrum

\item
2 Bl-Lac objects, 1 Compact Symmetric Object (CSO), and 1 Compact
Steep Spectrum Source (CSS).

\end{itemize}

Most of the extended FR I and FR II sources have a one-sided structure
on the pc-scale (17 FR I and 5 FR II; see Table 3).
We expect that sources at large angles to the line of sight, i.e.
those less affected by relativistic effects, show a two-sided 
structure, i.e. small ratios between the jet and the  counter-jet 
emission. We note, however, that the detection of a counter-jet
may be related 
to the sensitivity of the map, whereas the dynamic range is not 
a problem in our images, as the sources are generally weak. 
We classify as two-sided all sources showing both a jet and a counter-jet,
regardless of the value of the jet to counter-jet ratio, and of the
length of the counter-jet.
 
The total number of sources showing a two-sided morphology is 16,
corresponding to 30.2\%. This percentage is higher than that
found in previous samples in the literature: indeed there are 7/65
(11\%) symmetric sources in the Pearson-Readhead (PR) sample \citep{pe88},
and 19/411 (4.6\%) in the combined PR and Caltech-Jodrell
(CJ) samples \citep{tay94,po95}.
The difference in the percentage of two-sided jet sources can be even larger
if we consider that 
there are two separate classes of two-sided objects:
CSOs and core-jets. The CSOs are not strongly dependent on orientation
since most of the flux comes from the hot spots moving at 0.1 - 0.3 c.
Nearly all symmetric sources in the PR and CJ sample are
young CSO sources while very few of them are
expected in the local (z $<$ 0.1) universe.
We note that in our sample there is only one CSO 
\citep[4C31.04;][]{gir03} and it was not classified a two-sided jet source.
The difference between the percentage of symmetric sources in the
present sample and in previous samples is naturally explained in the
framework of unified scheme models by the fact that the present
sources show relatively faint cores, and are therefore less affected by
orientation bias.  
In particular, all two-sided FR II radio galaxies are
Narrow Line objects confirming that Broad Line Radio Galaxies are
oriented at least as close to the line-of-sight as quasars.

In a randomly oriented sample of radio galaxies,
the probability that a source is at an angle between $\theta_1$ and
$\theta_2$ with respect to the line-of-sight is:

$$ P(\theta_1,\theta_2) = cos(\theta_1) - cos(\theta_2)$$

\noindent
Thus, the percentage of FRI + FRII radio sources in the observed
sample (41/53 = 77\%) corresponds to sources oriented at angles $>$
40$^{\circ}$.  This is consistent with the fact that the observed
sample is likely representative of a sample of sources at random
angles to the line-of-sight.

It is difficult to estimate the expected number of sources showing
two-sided structure in our sample, since the detection of both the jet
and the counter-jet depends not only on the Doppler effect, but also
on the sensitivity needed to detect the counter-jet.  
The values of the jet to counter-jet ratios R for the
sources of our sample are given in Table 3.
The high values derived for 0055+30 (NGC 315) and 
0220+43 (3C 66B) are related to the presence of high brighntess 
jets and very deep radio observations of these sources \citep{g01}.
In the other two-sided sources, the values of R  are in  
the range 1 to 5. Most of the lower limits to R are much larger than 5,
thus confirming that the sensitivity in our maps
is generally good enough to detect as two-sided the sources
at large angles to the line of sight.
The sources with a two-sided structure in our sample are about 30\%.
This percentage implies that sources with two-sided jets
are oriented at angles $>$ 70$^\circ$.
This is consistent with a jet counter-jet ratio 
$\le$ 5 if  the jet velocity is of 0.9-0.99 c.

We note that 
a jet counter-jet ratio $\le$ 5 in two-sided sources implies also that
a counter-jet is easily detected only
if the jet brightness is at the level of about 15 r.m.s. (using a map
detection limit of 3 r.m.s.). 
We find that $\sim$ 37\% (10/27) of the FRI radio
galaxies and 44\% (4/9) of the FRII's are two-sided on the pc-scale. 
The higher fraction of symmetric sources among the FRII's could
suggest that the jets in FRII's are intrinsically stronger than those in
FRI's (assuming that the velocities are the same). 
We checked if the jet to counter-jet ratio is a function of the total
or arcsecond core radio power. We did not find any correlation while
as expected two sided sources are more frequently found in galaxies
with a low core dominance.
A more comprehensive analysis on
jet to counter-jet ratio will be performed when VLBI data on the full
radio galaxy sample are available.

Among the flat spectrum sources unresolved on the arcsecond scale, we
find 5 sources with a one-sided jet (including 2 BL-Lacs), 
and 2 sources with symmetric structures. 
2 other sources have not been detected, and 1 is
still unresolved on the pc-scale.  This large spread in pc-scale
structures suggests that this population could include intermediate low
power BL-Lac sources (one-sided), where relativistic effects are
important, as well as low-power sources whose intrinsically small size
might be related to a dense ISM and/or weak nuclear activity, unable
to power relativistic jets out to kpc-scales.

In most sources we find a good agreement between the pc and kpc-scale
structures. The comparison between the VLA and VLBA jet Position Angle
(PA - see Table 3) shows that only one source (1316+29) has a $\sim$
90$^\circ$ misalignment, while 2 other sources (1316+29; 1621+38) have
a difference of 50$^\circ$ and 30$^\circ$ respectively.  This result
supports the model where the large distortions detected in BL-Lac
sources are due to small intrinsic bends amplified by the small angle
of the BL-Lac jets to the line-of-sight. In all sources with the same
jet PA in the pc and kpc scale structure, the one sided pc scale
jet (or the brighter jet in double-sided pc scale jets) is oriented with 
respect to the nuclear emission, on the same side
of the brighter kpc scale jet (main jet) confirming the presence
of a continuous emission and not of a {\it flip-flop} jet
emission. Moreover this result suggests that the presence of 
a small jet asymmetry in the kpc scale images could be due to
Doppler boosting.

We have compared the correlated flux on the shortest VLBA baselines with
the core arcsecond flux density (Table 3). In 5 sources this
comparison was not possible because flux density variability or to the
presence of compact steep spectrum structures; among the remaining 48
sources, 34 (70\%) have a correlated flux density larger than 70\% of
the arcsecond core flux density.  Therefore in these sources we have
mapped most of the small scale structure and we are able to properly
connect the pc to the kpc structures. Finally for 14 (30\%) sources we
are missing a significant fraction of the arcsecond core flux density
(larger than 90\% in a few cases) in the VLBA images.  This result
suggests the presence of significant structures between $\sim$ 30 mas
and 1 arcsecond where VLBA can miss structures
because of the lack of short baselines and the VLA present angular resolution
can be too low.
To properly study these structures future observations with
the EVLA, the proposed New Mexico Array (EVLA phase 2) or the 
e-MERLIN array will be necessary.

\section {Conclusions}

In this paper we introduce a new complete sample of radio sources
selected from the B2 and 3CR catalogues, with z $<$0.1. For 53 of 95
sources we present VLBI data (28 of them for the first time) and
briefly discuss the pc and kpc-scale structure.

As expected, on the parsec scale a one-sided jet morphology is the
predominant structure present in our images, however $\sim$30\% of
sources observed so far show evidence of a two-sided structure which
has been quite rare in observations of sources observed to have
high power cores. This result is in agreement with a random orientation
of radio galaxies and a high jet velocity ($\beta$ $\sim$ 0.9).

With very few exceptions, the parsec and kpc-scale radio structures
are aligned confirming that the large bends present in some BL-Lac
sources are amplified by the small jet orientation angle with respect
to the line-of-sight. In sources with aligned pc and kpc scale structure
the main jet is  always on the same side with respect to the nuclear
emission. 

In the majority of the sources ($\sim$70\%) there is a good agreement
between the arcsecond scale and the VLBI core flux density. However for the
remaining 30\% of the sources the correlated VLBI flux density is
lower than 70\% of the arcsecond core flux density at the same
frequency.  This suggests the presence of a sub-kpc-scale structure
which will be best imaged with a new array such as the New Mexico Array.

No polarized flux has been detected confirming the low level of
polarized emission in radio galaxies at pc resolution as found by \citet{po03}.

A more detailed discussion on the properties of radio galaxies on the
pc-scale, such as jet intrinsic parameters, will be
presented in a future paper when data on the whole sample will 
be available.

\acknowledgments

We thank the referee for useful comments.
This work has received partial support by the Ministero dell'Istruzione,
dell'Universita' e della Ricerca (MIUR - Italy) under contract n. 2003027534. 

\vfill\eject

{}

\vfill\eject

\begin{deluxetable}{cccclrcc}
\tabletypesize{\small} 
\tablecaption{The Complete Bologna Sample\label{tab1}}
\tablehead{
\colhead{Name}&\colhead{Name}&\colhead{z}&\colhead{Morphology}
&\colhead{S$_c$(5.0)}&
\colhead{Log P$_c$}&\colhead{Log P$_t$}&\colhead{Notes}\\
\colhead{IAU}&\colhead{other}&\colhead{ }&\colhead{kpc}&\colhead{mJy}
&\colhead{W/Hz}&
\colhead{W/Hz}
&\colhead{ }}
\startdata

0034+25 & UGC367  & 0.0321 & FR I   &   10 & 22.65 & 24.08 & no VLBI \\
0055+26 & N326    & 0.0472 & FR I   &    8 & 22.89 & 25.68 & no VLBI \\
0055+30 & N315    & 0.0167 & FR I   &  588 & 23.85 & 24.56 & BSCS \\
0104+32 & 3C31    & 0.0169 & FR I   &   92 & 23.06 & 25.11 & BSCS \\
0106+13 & 3C33    & 0.0595 & FR II  &   24 & 23.58 & 26.69 & *       \\
0116+31 & 4C31.04 & 0.0592 & CSO    &   32 & 23.70 & 25.71 & BSCS \\
0120+33 & N507    & 0.0164 & FR I   &  1.4 & 21.21 & 23.91 & no VLBI \\
0149+35 & N708    & 0.0160 & FR I   &  5   & 21.74 & 23.60 & no VLBI \\ 
0206+35 & 4C35.03 & 0.0375 & FR I   &  106 & 23.82 & 25.46 & BSCS \\
0220+43 & 3C66B   & 0.0215 & FR I   &  182 & 23.56 & 25.59 & BSCS \\
0222+36 &         & 0.0327 &   C    &  140 & 23.76 & 24.20 & BSCS \\
0258+35 & N1167   & 0.0160 & CSS    &$<$243&$<$23.43 & 24.65 & BSCS \\
0300+16 & 3C76.1  & 0.0328 & FR I   & 10   & 22.67 & 25.38 & no VLBI \\
0326+39 &         & 0.0243 & FR I   &  78  & 23.30 & 24.69 & *       \\
0331+39 & 4C39.12 & 0.0202 &   C    &  149 & 23.42 & 24.49 & BSCS \\
0356+10 & 3C98    & 0.0306 & FR II  &   9  & 22.57 & 26.01 & no VLBI \\
0648+27 &         & 0.0409 &   C    &  213 & 24.19 & 24.31 & BSCS \\
0708+32 &         & 0.0672 & FR I   &   15 & 23.48 & 24.76 & no VLBI \\ 
0722+30 &         & 0.0191 & Spiral?&   51 & 22.39 & 24.42 &  no VLBI \\
0755+37 & N2484   & 0.0413 & FR I   &  195 & 24.16 & 25.65 & BSCS \\
0800+24 &         & 0.0433 & FR I   &    3 & 22.39 & 24.42 & no VLBI \\ 
0802+24 & 3C192   & 0.0597 & FR II  &    8 & 22.98 & 26.28 & no VLBI \\
0828+32AB&4C32.15 & 0.0507 & FR II  &   3.3& 22.57 & 25.69 & no VLBI \\
0836+29-II&4C29.30& 0.0790 & FR I   &  131 & 24.63 & 25.70 & BSCS \\
0836+29-I &4C29.30& 0.0650 & FR I   &  8.2 & 23.19 & 25.47 & no VLBI \\
0838+32 & 4C32.26 & 0.0680 & FR I   &  7.5 & 23.19 & 25.60 & no VLBI \\ 
0844+31 & IC2402  & 0.0675 & FR II  &  40  & 23.91 & 25.88 & * (1)   \\
0913+38 &         & 0.0711 & FR I   &$<$1.0&$<$22.35 & 25.27 & no VLBI \\
0915+32 &         & 0.0620 & FR I   &  8.0 & 23.13 & 24.90 & no VLBI \\
0924+30 &         & 0.0266 & FR I   &$<$0.4&$<$21.09&24.73 & no VLBI \\
1003+35 & 3C236   & 0.0989 & FR II  & 400  & 25.25 & 26.43 & * (1)   \\ 
1037+30 & 4C30.19 & 0.0909 &   C    &$<$84&$<$24.49 & 25.63& *       \\ 
1040+31 &         & 0.0360 &   C    & 55   & 23.49 & 24.98 & *       \\
1101+38 & Mkn 421 & 0.0300 & BL Lac & 640  & 24.40 & 24.66 & BSCS \\
1102+30 &         & 0.0720 & FR I   &  26  & 23.78 & 25.35 & *       \\
1113+29 & 4C29.41 & 0.0489 & FR I   &  41  & 23.64 & 25.12 & no VLBI \\
1116+28 &         & 0.0667 & FR I   &  30  & 23.77 & 25.33 & *       \\
1122+39 & N3665   & 0.0067 & FR I   &   6  & 21.07 & 22.75 & no VLBI \\ 
1142+20 & 3C264   & 0.0206 & FR I   & 200  & 23.57 & 25.46 & BSCS \\
1144+35 &         & 0.0630 & FR I   & 250  & 24.90 & 24.95 & BSCS \\
1204+24 &         & 0.0769 & FR I   &   8  & 23.32 & 24.85 & no VLBI \\ 
1204+34 &         & 0.0788 & FR II? &  23  & 23.80 & 25.45 & *       \\ 
1217+29 & N4278   & 0.0021 &   C    &$<$350&$<$21.08& 21.44& BSCS \\
1222+13 & 3C272.1 & 0.0037 & FR I   &  180 & 22.03 & 23.27 & BSCS \\
1228+12 & 3C274   & 0.0037 & FR I   & 4000 & 23.37 & 25.07 & BSCS \\
1243+26B&         & 0.0891 & FR I   &$<$1.8&$<$22.81 &25.50& no VLBI \\
1251+27 & 3C277.3 & 0.0857 & FR II  &   12 & 23.60 & 26.32 & no VLBI \\
1254+27 & N4839   & 0.0246 & FR I   &  1.5 & 21.60 & 23.74 & no VLBI \\
1256+28 & N4869   & 0.0224 & FR I   &   2.0& 21.64 & 24.50 & no VLBI \\
1257+28 & N4874   & 0.0239 & FR I   & 1.1  & 21.44 & 24.09 & no VLBI \\ 
1316+29 & 4C29.47 & 0.0728 & FR I   &  31  & 23.86 & 25.87 & *       \\ 
1319+42 & 3C285   & 0.0797 & FR I/II& 6    & 23.23 & 26.20 & no VLBI \\
1321+31 & N5127   & 0.0161 & FR I   & 21   & 22.37 & 24.61 & *       \\
1322+36 & N5141   & 0.0175 & FR I   & 150  & 23.30 & 24.36 & BSCS \\
1339+26 &         & 0.0722 & FR I   &$<$55&$<$22.11 & 25.47& no VLBI \\
1346+26 & 4C26.42 & 0.0633 & FR I   & 53   & 23.97 & 25.75 & *       \\
1347+28 &         & 0.0724 & FR I   & 4.8  & 23.05 & 25.08 & no VLBI \\
1350+31 & 3C293   & 0.0452 & FR I   &$<$100&$<$23.95 &25.96& *(1)    \\
1357+28 &         & 0.0629 & FR I   & 6.2  & 23.04 & 25.08 & no VLBI \\
1414+11 & 3C296   & 0.0237 & FR I   & 77   & 23.27 & 25.22 & *       \\
1422+26 &         & 0.0370 & FR I   & 25.0 & 23.18 & 25.08 & *       \\
1430+25 &         & 0.0813 & FR I   &$<$1.0&$<$22.47 &25.61& no VLBI \\ 
1441+26 &         & 0.0621 & FR I   &$<$0.7&$<$22.08 &25.05& no VLBI \\
1448+63 & 3C305   & 0.0410 & FR I   &  29  & 23.33 & 25.73 & *       \\
1502+26 & 3C310   & 0.0540 & FR I   & 80   & 24.01 & 26.48 & *       \\
1512+30 &         & 0.0931 & CSO?   &$<$0.4&$<$22.19 &25.02& no VLBI \\ 
1521+28 &         & 0.0825 & FR I   &  40  & 24.09 & 25.68 & *       \\
1525+29 &         & 0.0653 & FR I   & 2.5  & 22.68 & 24.91 & no VLBI \\
1528+29 &         & 0.0843 & FR I   & 4.5  & 23.16 & 25.37 & no VLBI \\
1529+24 & 3C321   & 0.0960 & FR II  & 30   & 24.10 & 26.47 & *       \\
1549+20 & 3C326   & 0.0895 & FR II  & 3.5  & 23.10 & 26.54 & no VLBI \\
1553+24 &         & 0.0426 & FR I   & 40   & 23.50 & 24.20 & *       \\
1557+26 & IC4587  & 0.0442 &   C    & 31   & 23.43 & 23.82 & *       \\ 
1610+29 & N6086   & 0.0313 & FR I   &$<$6.0&$<$22.41 &24.14& no VLBI \\
1613+27 &         & 0.0647 & FR I   & 25   & 23.67 & 24.77 & *       \\ 
1615+35 & N6107   & 0.0296 & FR I   & 28   & 23.03 & 25.31 & no VLBI \\ 
1621+38 & N6137   & 0.0310 & FR I   & 50   & 23.32 & 24.59 & *       \\ 
1626+39 & 3C338   & 0.0303 & FR I   & 105  & 23.62 & 25.86 & BSCS \\
1637+29 &         & 0.0875 & FR I   & 13   & 23.65 & 25.44 & no VLBI \\ 
1652+39 & Mkn 501 & 0.0337 & BLLac  & 1250 & 24.79 & 24.96 & BSCS \\
1657+32A&         & 0.0631 & FR I   & 2.5  & 22.64 & 25.34 & no VLBI \\ 
1658+30 & 4C30.31 & 0.0351 & FR I   & 84   & 23.66 & 24.92 & *       \\
1736+32 &         & 0.0741 & FR I   &  8   & 23.29 & 25.17 & no VLBI \\ 
1752+32B&         & 0.0449 & FR I   & 12   & 23.03 & 24.45 & no VLBI \\ 
1827+32A&         & 0.0659 & FR I   & 26   & 23.70 & 25.15 & *       \\
1833+32 & 3C382   & 0.0586 & FR II  & 188  & 24.46 & 26.31 & BSCS \\
1842+45 & 3C388   & 0.0917 & FR II  &  62  & 24.37 & 26.74 & *(1)     \\
1845+79 & 3C390.3 & 0.0569 & FR II  & 330  & 24.67 & 26.58 & BSCS \\
1855+37 &         & 0.0552 &   C    &$<$100&$<$24.13 &25.03& *       \\
2116+26 & N7052   & 0.0164 & FR I   & 47   & 22.74 & 23.57 & Xu et al.(2000)\\
2212+13 & 3C442   & 0.0262 & FR I?  & 2.0  & 21.76 & 25.50 & no VLBI \\
2229+39 & 3C449   & 0.0181 & FR I   & 37   & 22.72 & 24.96 & *       \\
2236+35 &         & 0.0277 & FR I   & 8.0  & 22.43 & 24.41 & no VLBI \\
2243+39 & 3C452   & 0.0811 & FR II  & 130  & 24.58 & 26.92 & BSCS \\
2335+26 & 3C465   & 0.0301 & FR I   & 270  & 23.99 & 25.91 & BSCS \\

\enddata
\tablecomments{ S$_c$(5.0) is the arc-second core flux density at 5.0 GHz and 
Log P$_c$ is the corresponding logarithm of the radio power;
Log P$_t$ is the logarithm of the total radio power at 408 MHz; Morphology:
C = flat spectrum compact core, CSO = Compact Symmetric Object, 
CSS = Compact Steep Spectrum source, FR-I or II = Fanaroff type I or II; 
Notes refer to the status of VLBI observations:
BSCS = see Giovannini et al. 2001; * new data are given in this paper;
(1) = re-started source.
For a reference to the extended structure see Fanti et al. 1987 for B2 sources
and Leahy et al. HTML page:
http://www.jb.man.ac.uk/atlas).}
\end{deluxetable}

\vfill\eject

\begin{deluxetable}{ccrrrccc}
\tabletypesize{\small} 
\tablecaption{Observed Source Parameters\label{tab2}}
\tablehead{
\colhead{Name}&\colhead{Name}&\colhead{S$_c$(5.0)}&\colhead{S$_{VLBI}$(tot)}&
\colhead{S$_{VLBI}$(core)}&\colhead{HPBW}&\colhead{Noise}&
\colhead{Polarization}\\
\colhead{IAU}&\colhead{other}&\colhead{mJy}&\colhead{mJy}&
\colhead{mJy}&\colhead{mas}&\colhead{mJy/beam}&\colhead{\%}}
\startdata

0106+13 & 3C33    & 24 & 30 & 11.0 & 3.5$\times$1.4 (-5) & 0.12& $<$ 3.0  \\
0326+39 &         & 78 & 74 & 55.5 & 2.8$\times$1.6 (11) & 0.14& $<$ 1.5   \\
0844+31 & IC2402  & 40 & 32 & 27.0 & 3.4$\times$2.0 (19) & 0.08& $<$ 1.5  \\
1003+35 & 3C236   & 400 &466 &152.8 & 2.9$\times$2.0 (25) & 0.10& $<$ 0.2  \\
1037+30 &         &$<$ 84&$<$ 2&N.D.& 3.4$\times$2.0 (15)&  0.15&  -  \\
1040+31 &         & 55 & 48 & 37.6 & 2.7$\times$2.1 (23) & 0.17& $<$ 1.0    \\
1102+30 &         & 10 & 17 & 14.3 & 2.5$\times$1.6 (-6) & 0.14& $<$ 2.5    \\
1116+28 &         & 30 & 11 & 9.2  & 2.9$\times$1.7 (-5) & 0.14& $<$ 3.5   \\
1204+34 &         & 23 & 39 & 33.7 & 2.7$\times$1.7 (-3) & 0.16& $<$ 1.0  \\ 
1316+29 & 4C29.47 & 31 & 23 &  5.3 & 2.5$\times$2.5       & 0.09& $<$ 6.0  \\ 
1321+31 & N5127   & 21 & 11 &  3.7 & 2.8$\times$1.7 (-7)  & 0.12& $<$ 10.0  \\
1346+26 & 4C26.42 & 53 & 8  &  5.8 & 3$\times$3           & 0.10& $<$ 5.0  \\
1350+31 & 3C293   &100 & 23 & 14.4 & 3$\times$2.5 (0)     & 0.10& $<$ 2.0 \\
1414+11 & 3C296   & 77 & 86 & 65.1 & 3.6$\times$1.7 (-4)  & 0.09& $<$ 0.4   \\
1422+26 &         & 25 & 3  & 1.7  & 3.5$\times$2.0 (-2)  & 0.05&  -    \\
1448+63 & 3C305   & 29 & 2  & 1.1  & 3.6$\times$2.0 (-5)  & 0.06&  -  \\
1502+26 & 3C310   & 88 & 3  & 2.7  & 3.6$\times$2.2 (-3)  & 0.06&  -  \\
1521+28 &         & 56 & 44 & 39.5 & 2.9$\times$1.6 (1)    & 0.04& $<$ 0.3  \\
1529+24 & 3C321   & 30 & 3  & 3.4  & 3.5$\times$2.1 (3)    & 0.05&   -  \\
1553+24 &         & 40 & 46 & 40.7 & 3.5$\times$1.9 (-1)   & 0.05& $<$ 0.4   \\
1557+26 & IC4587  & 31 & 9  & 7.8  & 3.2$\times$1.6 (-19)  & 0.26& $<$ 10.0 \\
1613+27 &         & 25 & 8  & 6.8  & 3.2$\times$1.6 (-19)  & 0.24&  -  \\
1621+38 & N6137   & 50 & 20 & 12.9 & 3.0$\times$1.6 (-24)  & 0.11& $<$ 2.5 \\ 
1658+30 & 4C30.31 & 84 & 76 & 60.0 & 3.2$\times$1.5 (-17)  & 0.14& $<$ 0.7  \\
1827+32A&         & 26 & 15 & 5.9  & 2.9$\times$1.7 (-25)  & 0.12& $<$ 10.0  \\
1842+45 & 3C388   & 62 & 52 & 34.0 & 2.6$\times$1.7 (-27)  & 0.15& $<$ 1.0 \\
1855+37 &         &$<$34&$<$2&N.D. & 3.0$\times$1.5 (-25)  & 0.11& -   \\
2229+39 & 3C449   & 37 & 30 & 25.9 & 3.0$\times$1.9 (-23)  & 0.12& $<$ 1.5 \\

\enddata
\tablecomments{S$_c$ is the arcsecond core flux density at 5 GHz; S$_{VLBI}$(tot)
is the VLBI correlated flux in the shortest baselines; S$_{VLBI}$(core) is the nuclear
source flux density in the VLBI images.}
\end{deluxetable}

\vfill\eject

\begin{deluxetable}{ccccccc}
\tabletypesize{\small} 
\tablecaption{Parsec and Kiloparsec-scale Comparison}
\tablehead{
\colhead{Name}&\colhead{Name}&
\colhead{Morphology}&\colhead{j/cj ratio}&\colhead{Morphology}&
\colhead{S$_{VLBI}/S_{VLA}$}&\colhead{$\Delta$PA}\\
\colhead{IAU}&\colhead{other}&\colhead{kpc}&\colhead{}&
\colhead{pc}&\colhead{\%}&\colhead{degree}}

\startdata
0055+30 & N315    & FR I  &   42 & 2s&  60& 0  \\
0104+32 & 3C31    & FR I  &$>$16 & 1s& 100& 5 \\
0106+13 & 3C33    & FR II &  2.0 & 2s& 100& 0 \\
0116+31 & 4C31.04 & CSO   &   -  & CSO& 100& 0 \\
0206+35 & 4C35.03 & FR I  &$>$14 & 1s&  80& 0 \\
0220+43 & 3C66B   & FR I  &   10 & 2s& 100& 0 \\
0222+36 &         & C     &   -  & C & 100& 0  \\
0258+35 & N1167   & CSS   &   -  & CSS&  ? & ?  \\
0326+39 &         & FR I  &$>$25 &  1s& 100& 0  \\
0331+39 & 4C39.12 & C     &$>$12 & 1s&  74& 0  \\
0648+27 &         & C     &$>$7.5& 1s&  85& -  \\
0755+37 & N2484   & FR I  &$>$20 & 1s& 100& 0  \\
0836+29-II&4C29.3 & FR I  &$>$20 & 1s& 100& 0  \\
0844+31 & IC2402  & FR II &  1.0 & 2s&  80& 0  \\
1003+35 & 3C236   & FR II &$>$45 & 1s&   100 & 0  \\
1037+30 & 4C30.19 & C     &   -  & ND& - & -  \\ 
1040+31 &         & C     &    5 & 2s&  80& 0  \\
1101+38 & Mkn 421 & C     &$>$110& 1s&  70& 0  \\
1102+30 &         & FR I  &$\sim$1.7?& 2s?& 100& 0  \\
1116+28 &         & FR I  &   3.3& 2s&  40& 0  \\
1142+20 & 3C264   & FR I  &$>$24 & 1s&  60& 0  \\
1144+35 &         & FR I  &see note & 2s&  - & 0  \\
1204+34 &         & FR II &$>$2.0& 1s& 100& 0 \\ 
1217+29 & N4278   & C     &   2.3& 2s& 100& - \\
1222+13 & 3C272.1 & FR I  &$>$10 & 1s&  70 & 0  \\
1228+12 & 3C274   & FR I  &$>$150& 1s&  25& 0  \\
1316+29 & 4C29.47 & FR I  &   1.0& 2s&  70&50 \\ 
1321+31 & N5127   & FR I  & 1?   & 2s?&  50& 0 \\
1322+36 & N5141   & FR I  &$>$20 & 1s&  40& 0 \\
1346+26 & 4C26.42 & FR I  &  -   & C? &  15& - \\
1350+31 & 3C293   & FR II &  2.0 & 2s&  20& 0 \\
1414+11 & 3C296   & FR I  &  2.0 & 2s& 100& 0 \\
1422+26 &         & FR I  &$>$20?&1s?&  10& 0  \\
1448+63 & 3C305   & FR I  &   - & C?&   7& 90 \\
1502+26 & 3C310   & FR I  &$>$2.0& 1s&   5& 0  \\
1521+28 &         & FR I  &$>$17 & 1s&  80& 0 \\
1529+24 & 3C321   & FR II &   -  & C & 10 & -  \\
1553+24 &         & FR I  &$>$20 & 1s& 100& 0  \\
1557+26 & IC4587  & C     &$>$ 4 & 1s&  30& -  \\
1613+27 &         & FR I  &   -  & C  &  30& -  \\
1621+38 & N6137   & FR I  &$>$ 5 & 1s&  40& 30  \\ 
1626+39 & 3C338   & FR I  &  1.1 &  2s&  - & 0  \\
1652+39 & Mkn 501 & BLLac &$>$1250& 1s&  80& -  \\
1658+30 & 4C30.31 & FR I  &$>$8.5& 1s&  90& 0  \\
1827+32A&         & FR I  &$>$7.5& 1s&  60& 0 \\
1833+32 & 3C382   & FR II &$>$20 & 1s&  70& 0  \\
1842+45 & 3C388   & FR II &$>$10 & 1s&  80& 0  \\
1845+79 & 3C390.3 & FR II &$>$100& 1s&  ? & 0  \\
1855+37 &         & C     &   -  & ND& -  & - \\
2116+26 & N7052   & FR I  & 1.0? & 2s?&  - & 0  \\
2229+39 & 3C449   & FR I  &   -  & C & 80 & -  \\
2243+39 & 3C452   & FR II & 1.1 & 2s & 100& 15 \\
2335+26 & 3C465   & FR I  &$>$10 & 1s & 100& 0  \\

\enddata
\tablecomments{1144+35 is a two-sided source with a large
flux density variability in the jet structure, see \citet{g99}.
In col. 6  
is given the ratio between the arcsecond core flux density and the 
VLBA correlated flux (at 5 GHz). $\Delta$PA (col. 7)
is the difference between the 
kpc and pc 
scale jet PA}
\end{deluxetable}

\clearpage

\begin{figure}
\epsscale{0.3}
\plotone{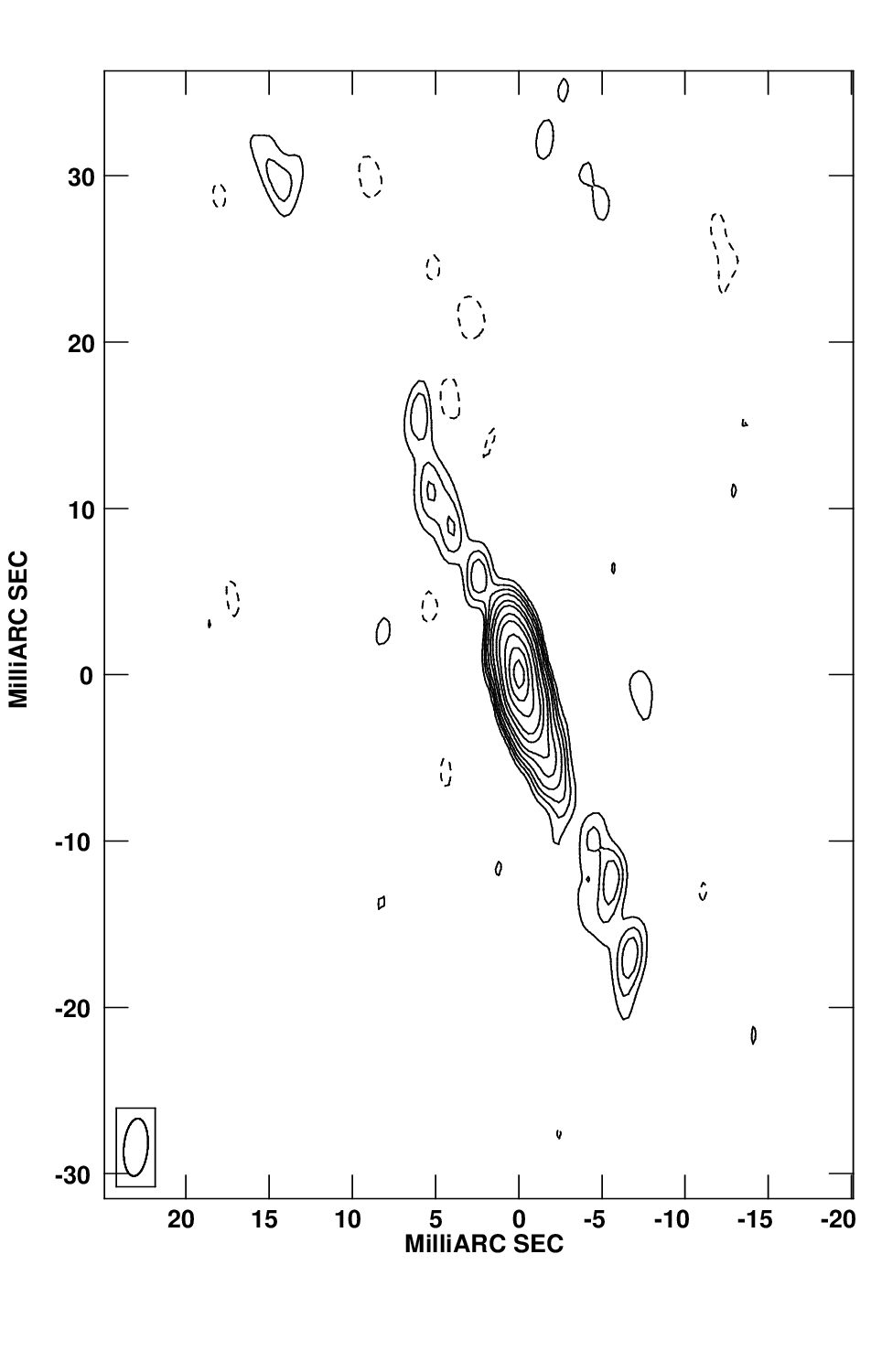}
\caption{VLBA image of the narrow line FR II galaxy 0106+13 (3C33), at 5 GHz.
Levs are: $-$0.3 0.3 0.5 0.7 1 1.5 2 3 4 6 8 10 mJy/beam.  The restoring
beams of the images presented in this figure and in the following
figures are given in Tab. 2}
\end{figure}

\begin{figure}
\epsscale{0.4}
\plotone{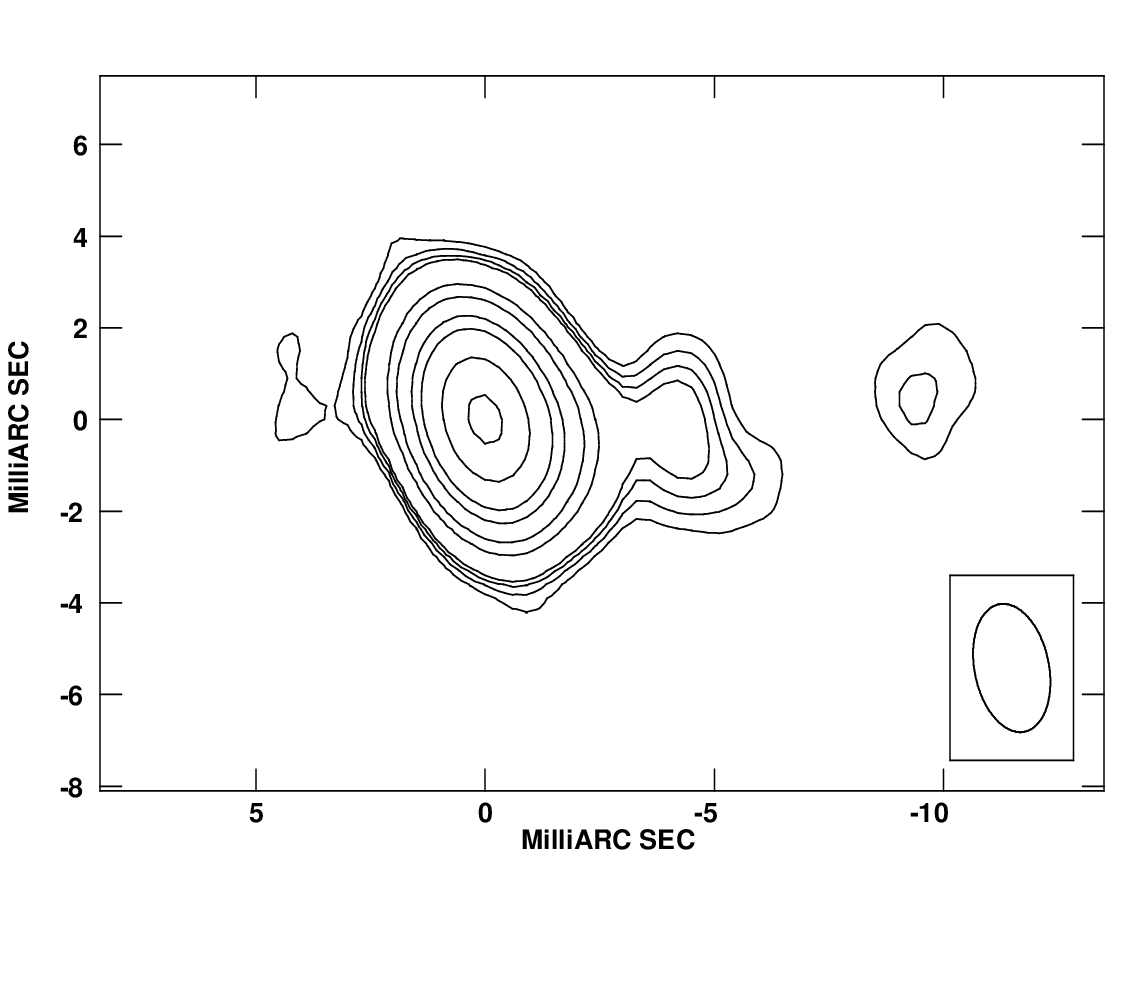}
\caption{VLBA image at 5 GHz of 0326+39. Levs are: $-$0.4 0.4 0.6 0.8 1 3 5 10 
15 30 50 mJy/beam}
\end{figure}

\begin{figure}
\epsscale{0.7}
\plottwo{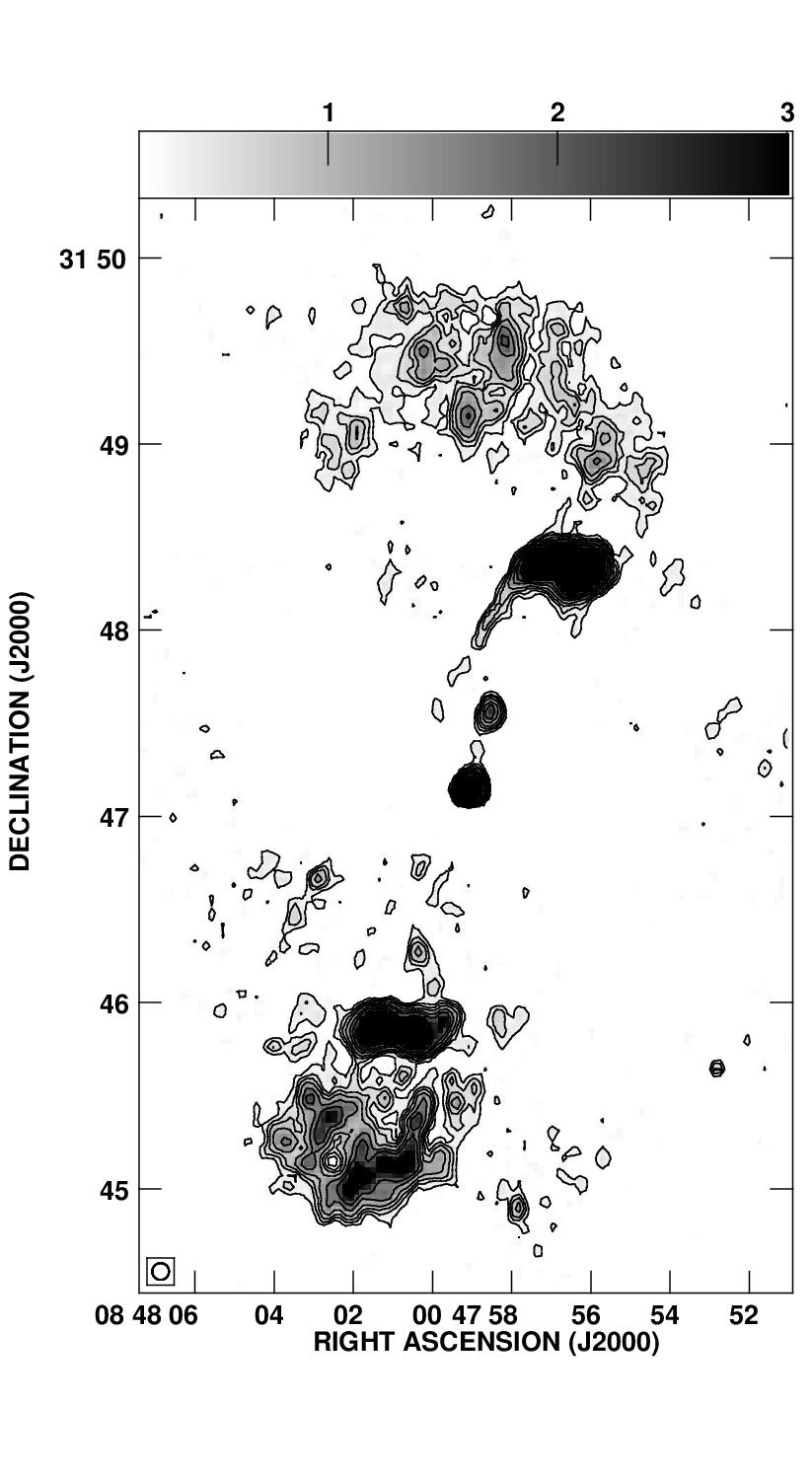}{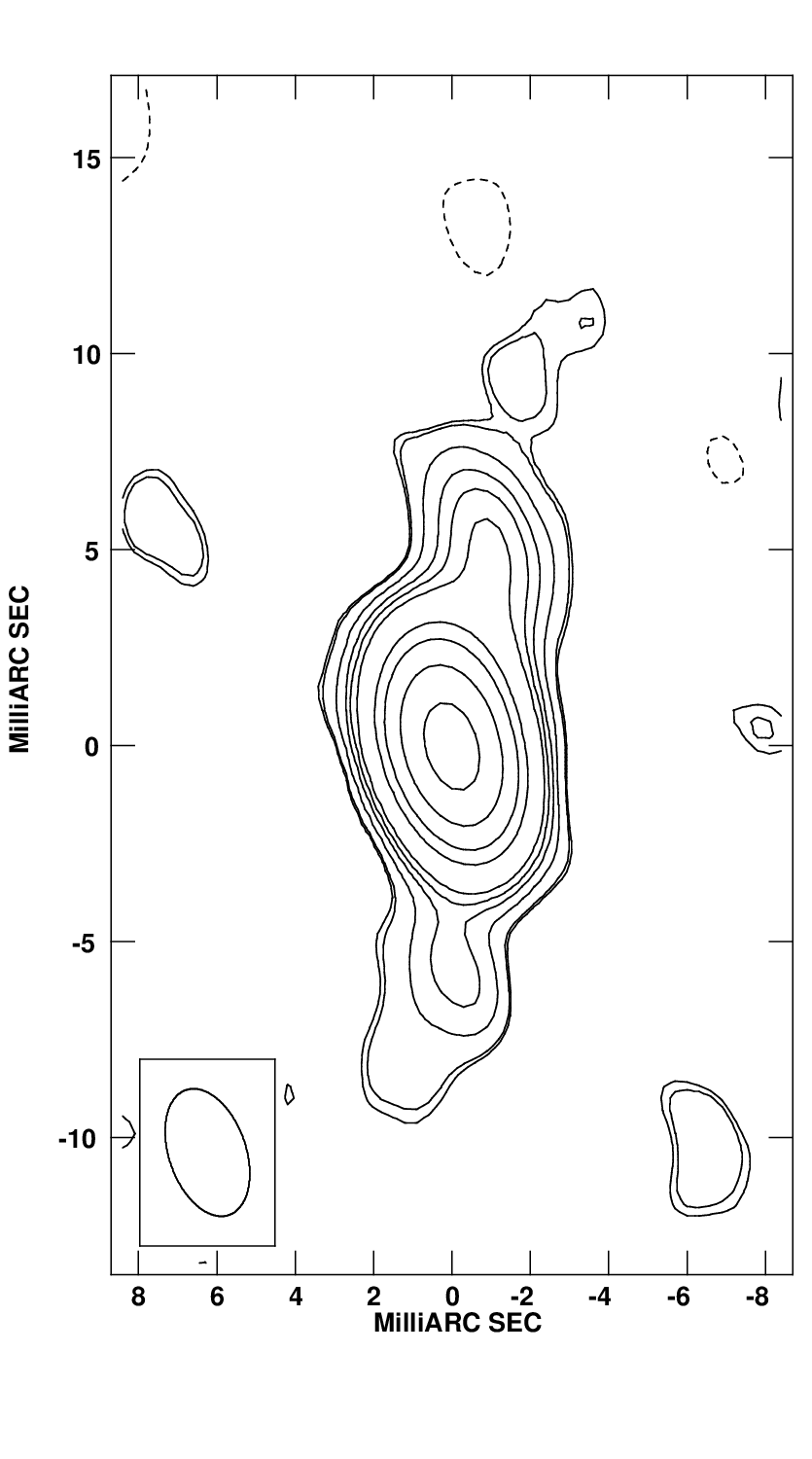}
\caption{VLA image from the FIRST survey (left) of 0844+31 (IC2402)with levs =
0.3 0.5 0.7 1 1.5 2 3 5 7 10 20 30 mJy/beam (the HPBW is 
5$^{\prime\prime}$). VLBA image at 5 GHz of 0844+31 (right);
levs are: -0.15 0.13 0.15 0.3 0.5 0.7 1 3 5 10 20 mJy/beam.}
\end{figure}

\begin{figure}
\epsscale{0.9}
\plottwo{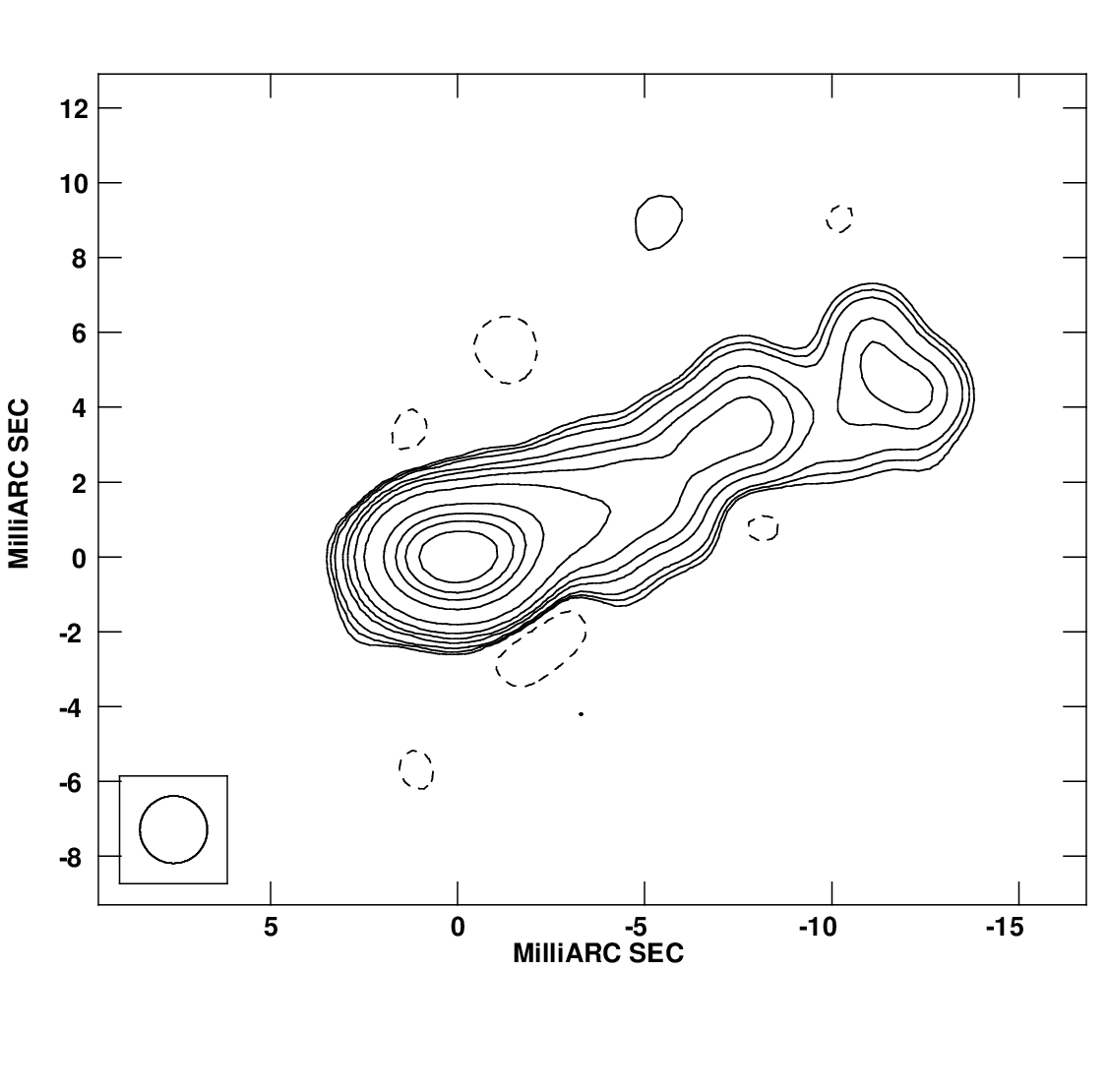}{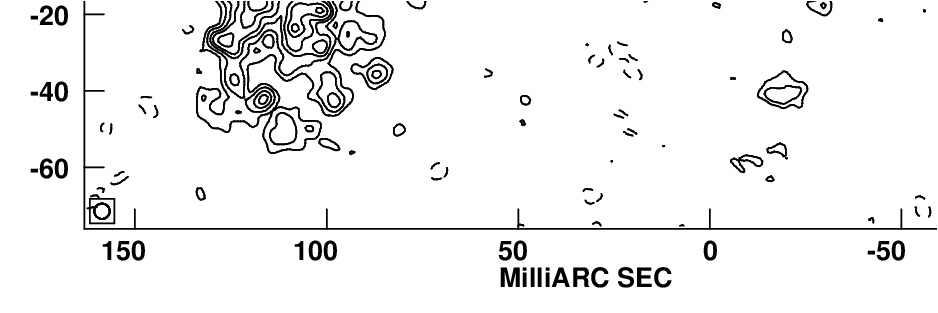}
\caption{VLBA image at 5 GHz of 1003+35 (3C236) at high (left) and low 
resolution (right.
 Levs are: $-$0.5 0.5 0.7 1 2 3 5 10 30 50 70 100 mJy/beam (left) and
$-$0.2 0.2 0.4 0.7 1 1.5 2 3 5 10 20 50 70 100 150 200 mJy/beam (right). }
\end{figure}

\begin{figure}
\epsscale{0.3}
\plotone{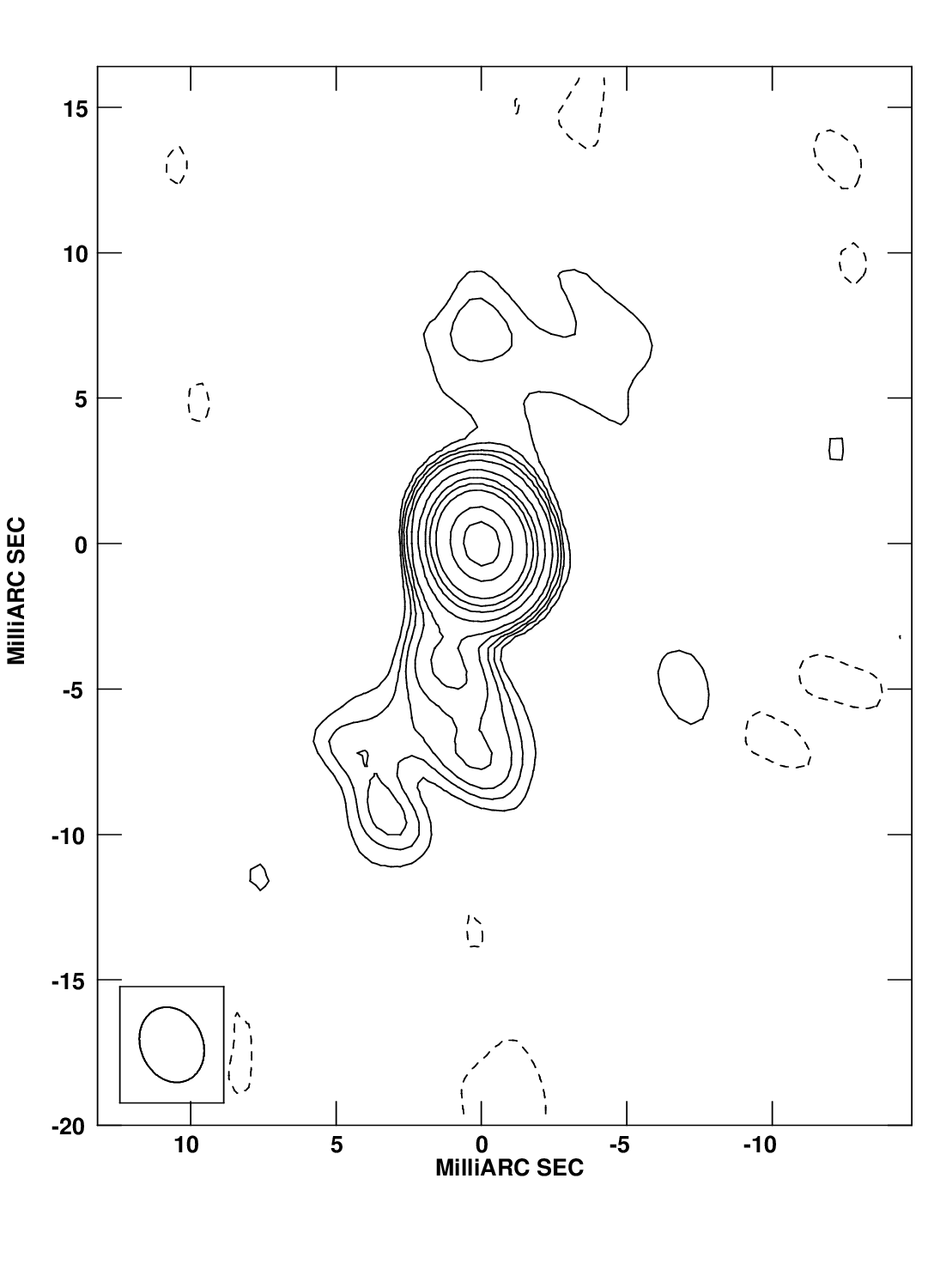}
\caption{VLBA image at 5 GHz of 1040+31. Levs are: $-$0.3 0.3 0.5 0.7 1 1.5 3 5 7 10 20 30 mJy/beam.}
\end{figure}

\begin{figure}
\epsscale{0.9}
\plottwo{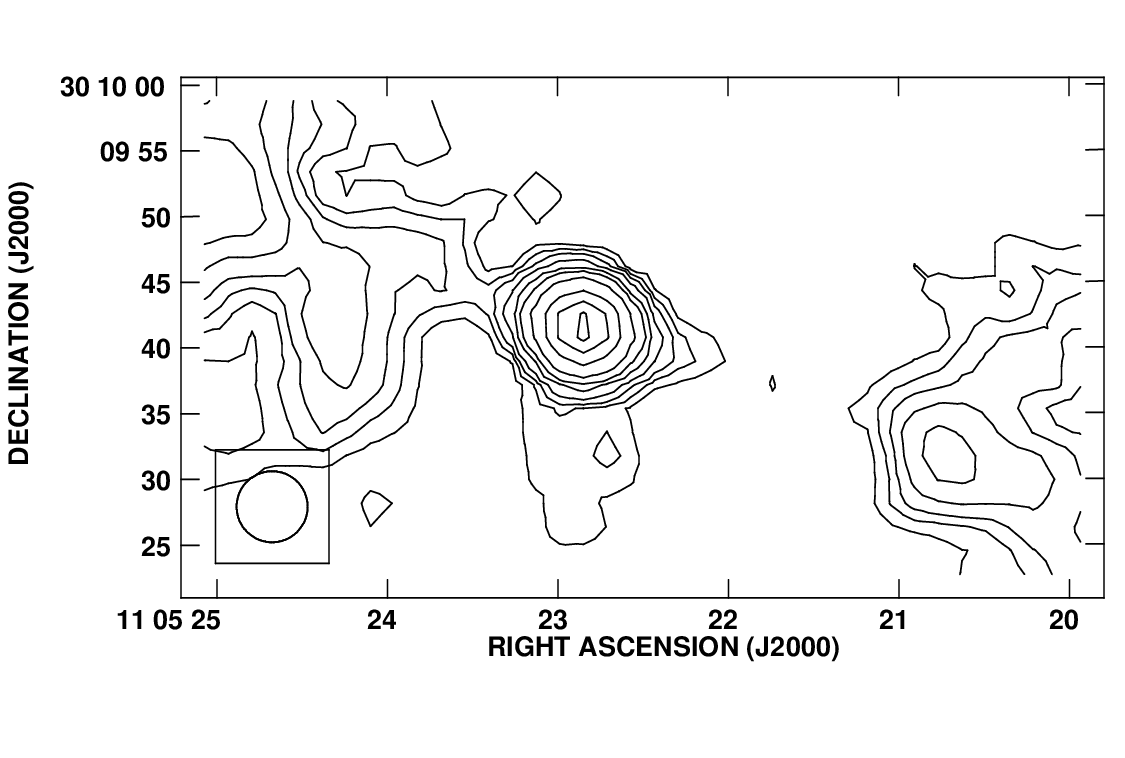}{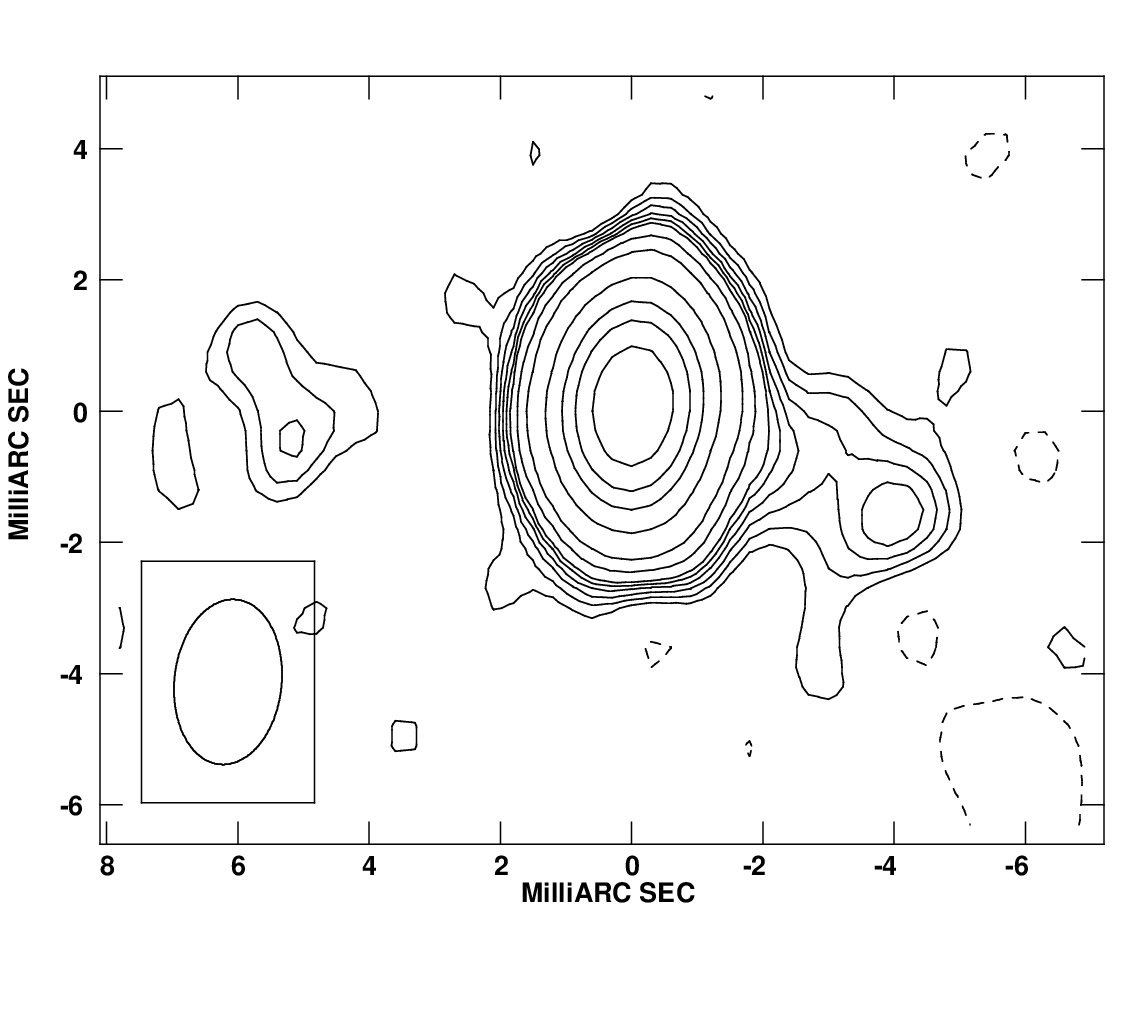}
\caption{VLA image from the FIRST survey (left) of 1102+30, with
5$^{\prime\prime}$ resolution. Only the innermost source region is
displayed to show the direction of the main jet (West) at the
arcsecond resolution.  Levs are 0.3 0.5 0.7 1 1.5 2 3 5 7 9 mJy/beam;
VLBA image at 5 GHz of 1102+30 (right) with Levs = $-$0.2 0.2 0.3 0.4
0.5 0.6 0.7 1 1.5 3 5 7 10 mJy/beam.}
\end{figure}

\begin{figure}
\epsscale{0.5}
\plotone{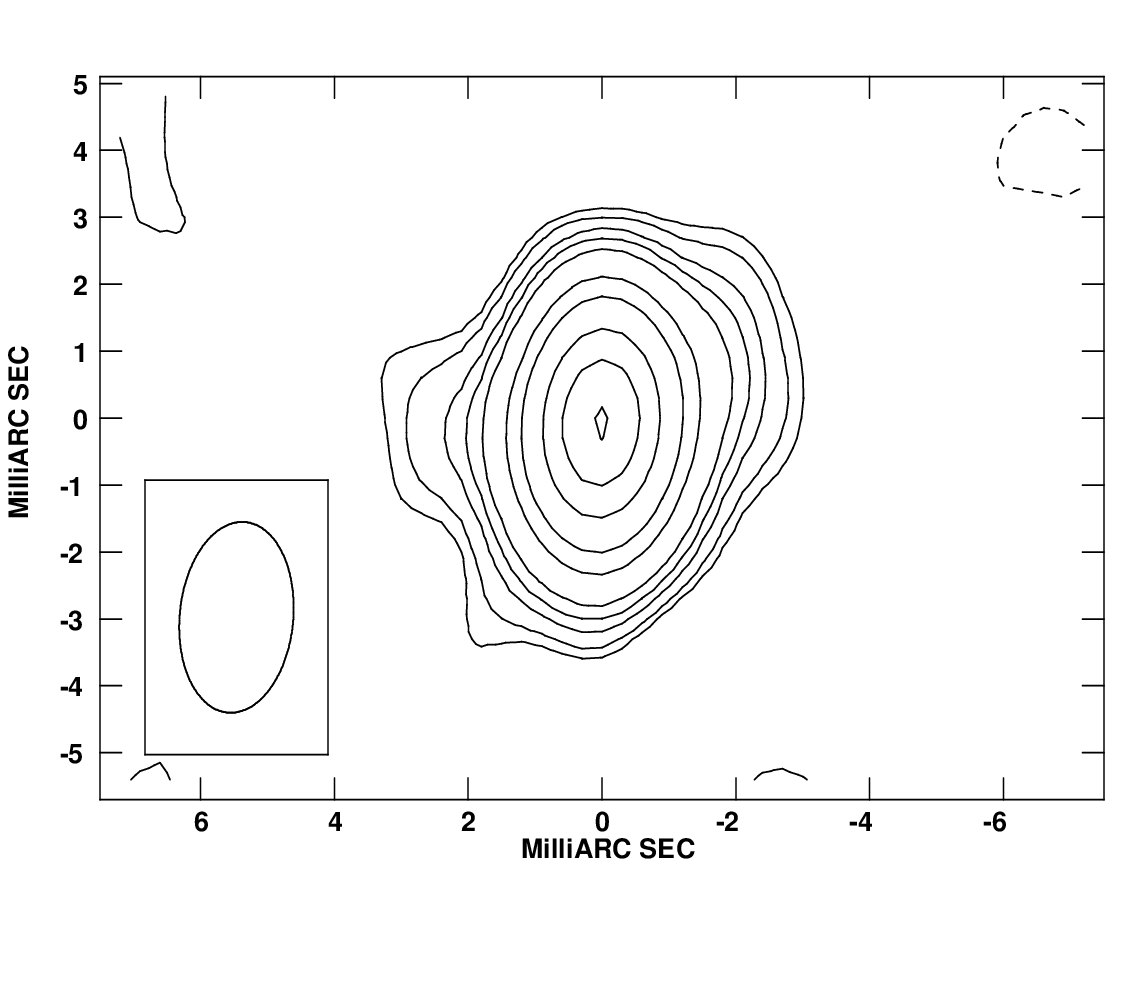}
\caption{VLBA image at 5 GHz of 1116+28. Levs are: $-$0.3 0.2 0.3 0.5 0.7 1 2 3 
5 7 9 mJy/beam.}
\end{figure}

\begin{figure}
\epsscale{0.5}
\plotone{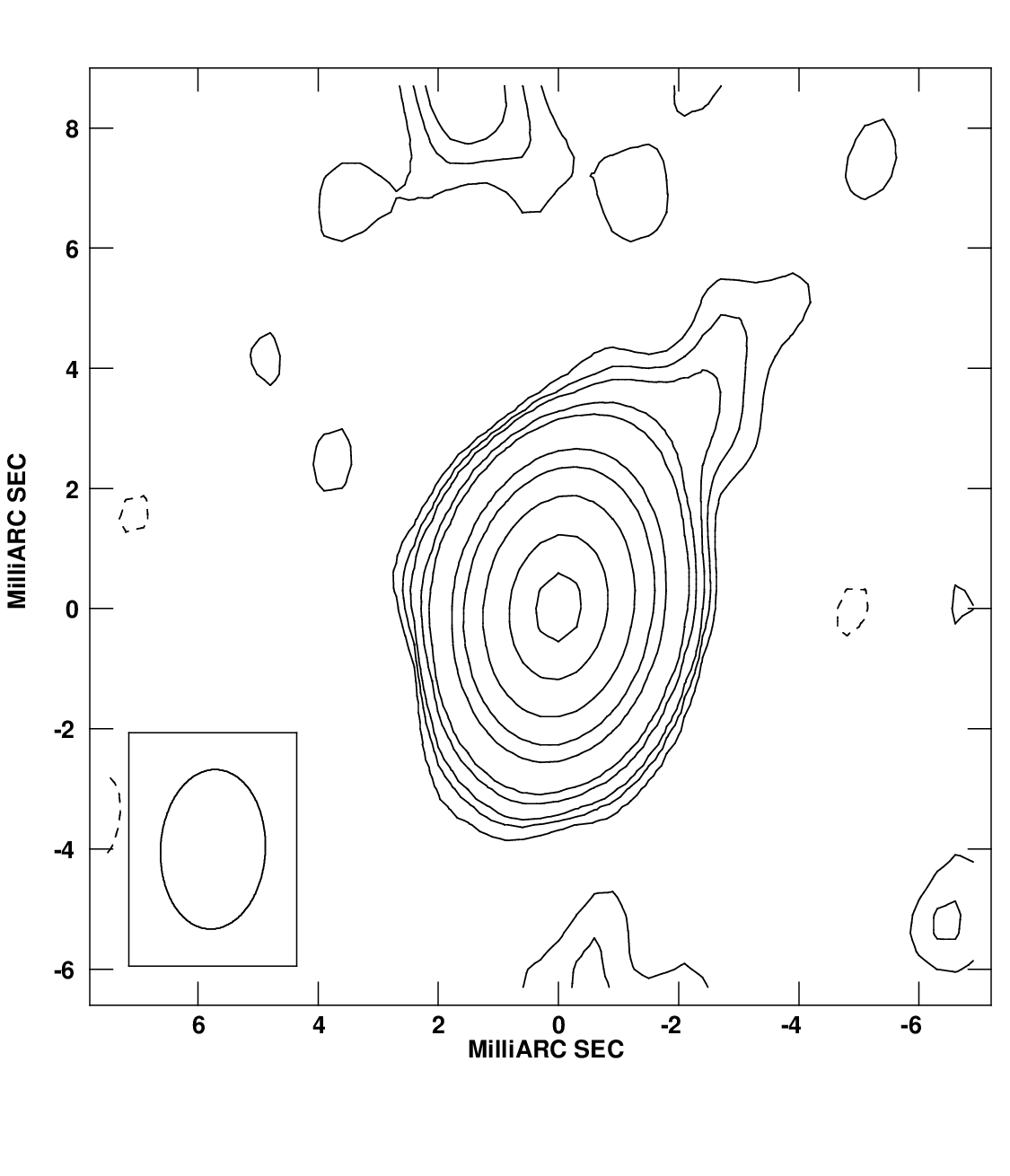}
\caption{VLBA image at 5 GHz of 1204+34. Levs are: $-$0.3 0.2 0.3 0.4 0.7 1 3 5 
10 20 30 mJy/beam}
\end{figure}

\begin{figure}
\epsscale{0.5}
\plotone{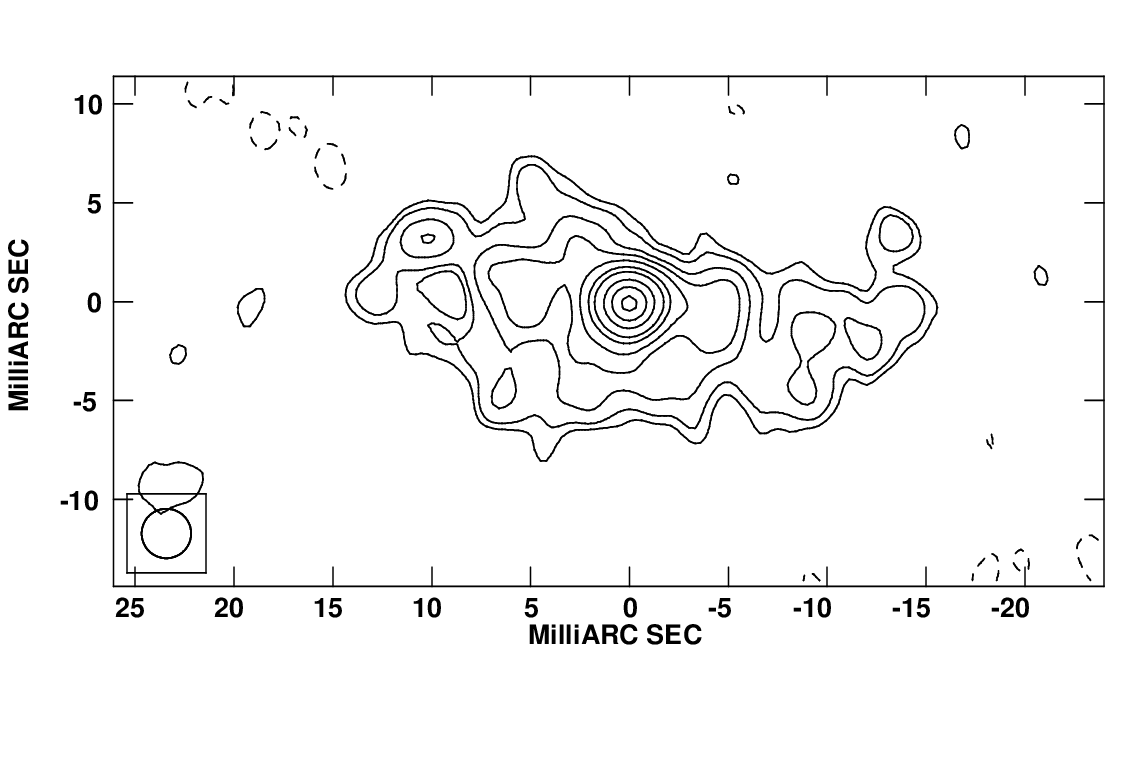}
\caption{VLBA image at 5 GHz of 1316+29 (4C29.47). 
Levs are: $-$0.2 0.2 0.3 0.5 0.7 1 1.5 
2 3 4 5 mJy/beam.}
\end{figure}

\begin{figure}
\epsscale{0.5}
\plotone{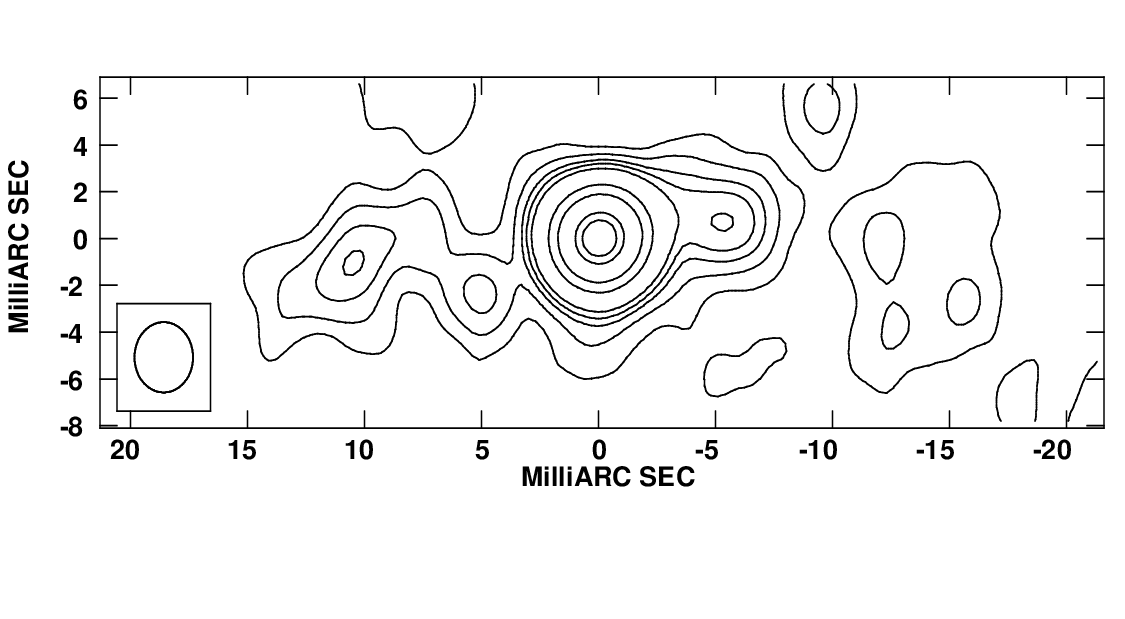}
\caption{VLBA image at 5 GHz of 1350+31 (3C 293). 
Levs are: $-$0.15 0.15 0.3 0.5 0.7 1 3 
5 10 12 mJy/beam.}
\end{figure}

\begin{figure}
\epsscale{0.4}
\plotone{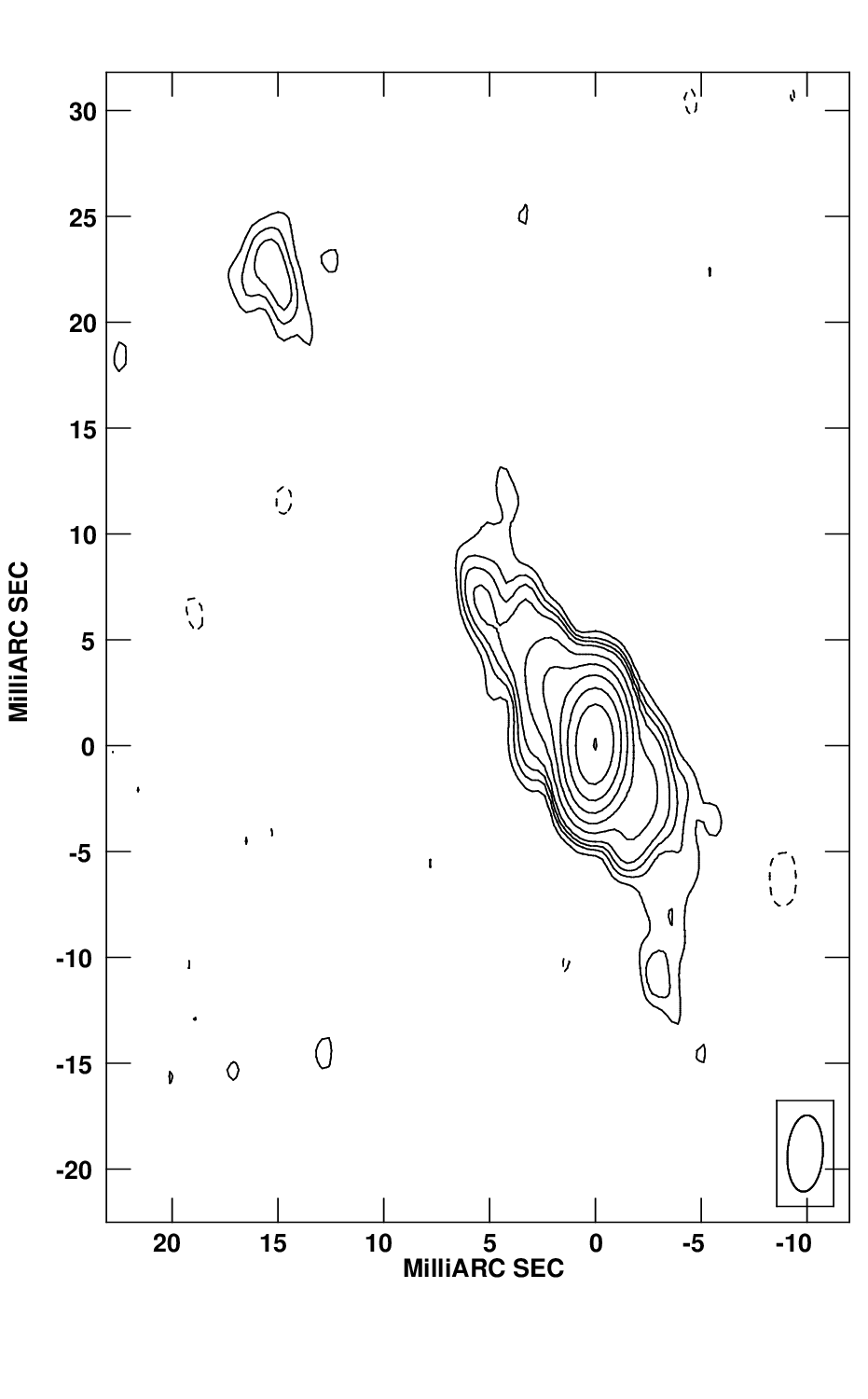}
\caption{VLBA image at 5 GHz of 1414+11 (3C 296). 
Levs are: $-$0.3 0.3 0.5 0.7 1 2 4 8 16 
32 64 mJy/beam.}
\end{figure}

\begin{figure}
\epsscale{0.4}
\plotone{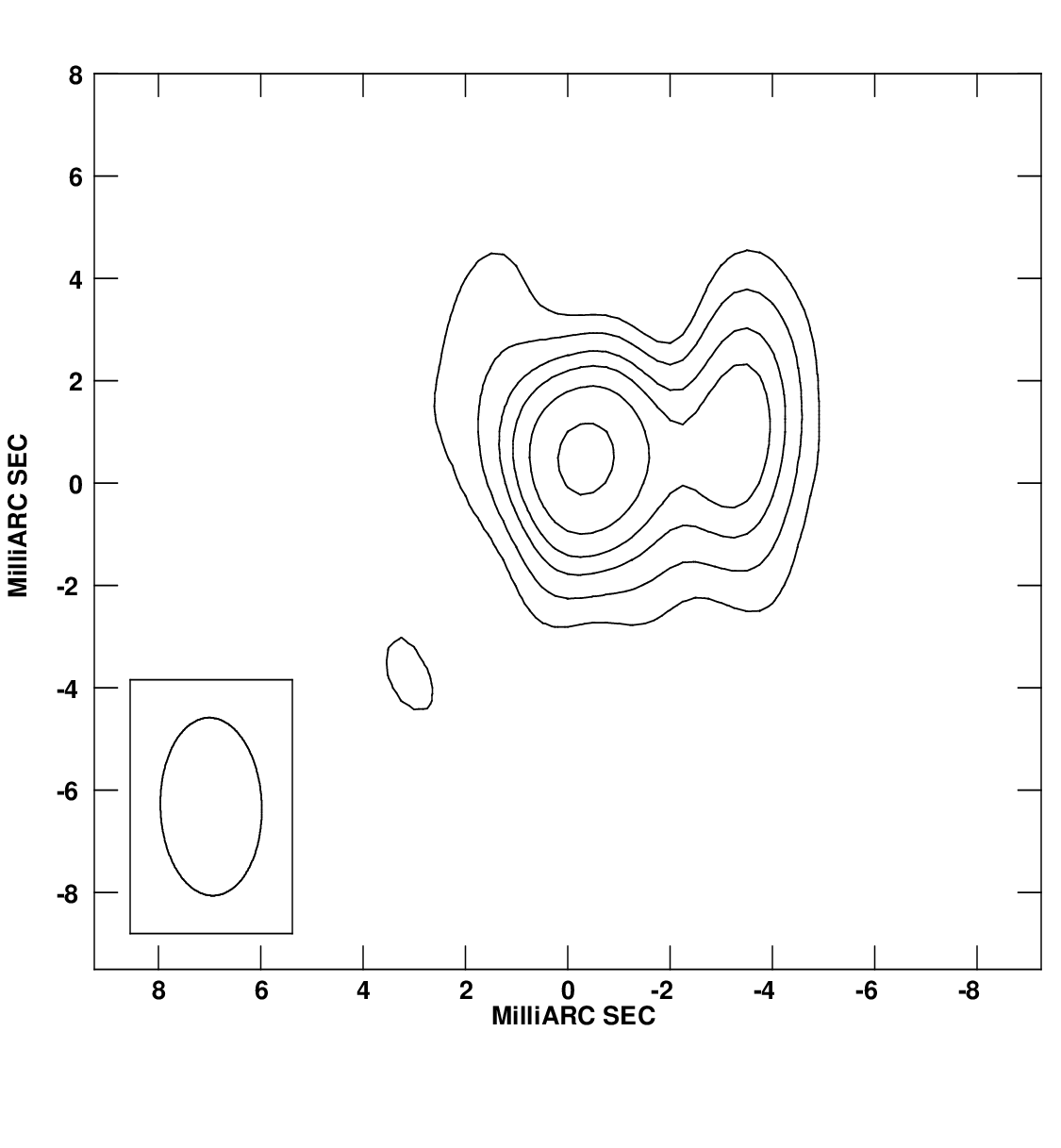}
\caption{VLBA image at 5 GHz of 1422+26. Levs are: $-$0.15 0.15 0.3 0.5 0.7 1.5 
mJy/beam}
\end{figure}

\begin{figure}
\epsscale{0.4}
\plotone{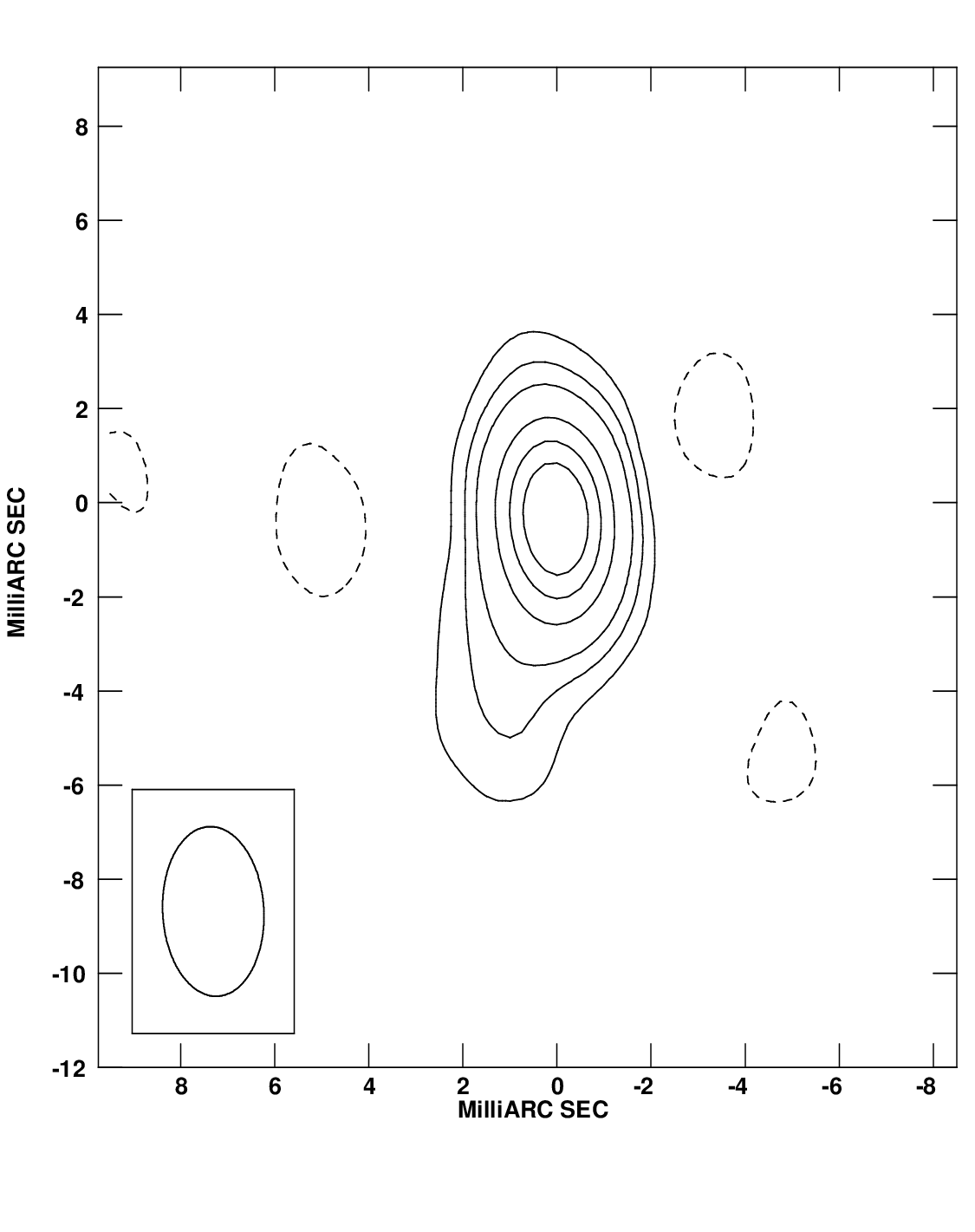}
\caption{VLBA image at 5 GHz of 1502+26 (3C 310). 
Levs are: $-$0.15 0.15 0.3 0.5 1 1.5 2
mJy/beam}
\end{figure}

\begin{figure}
\epsscale{0.4}
\plotone{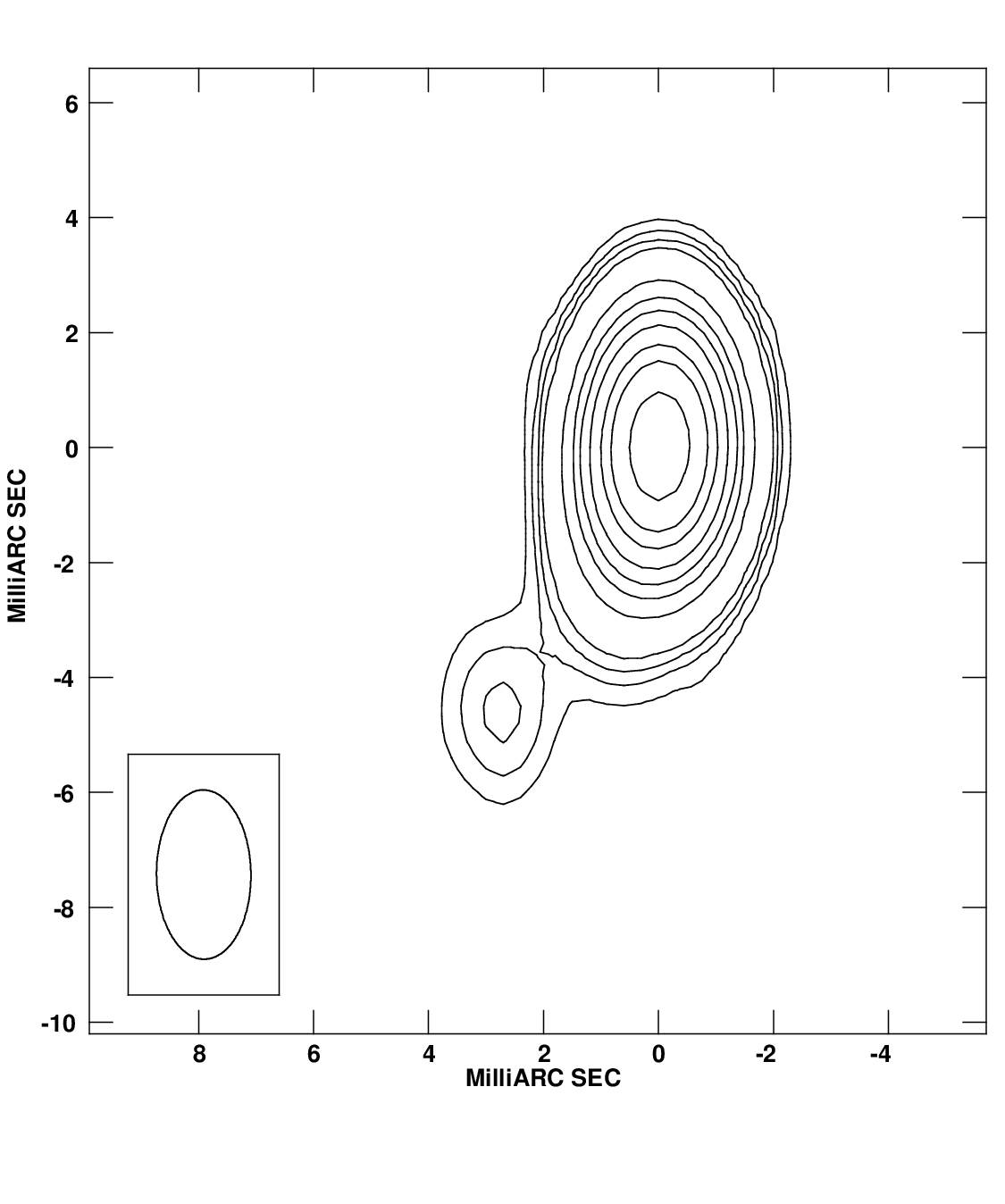}
\caption{VLBA image at 5 GHz of 1521+28. Levs are: $-$0.3 0.3 0.5 0.7 1 3 5 7 
10 15 20 30 mJy/beam.}
\end{figure}

\begin{figure}
\epsscale{0.4}
\plotone{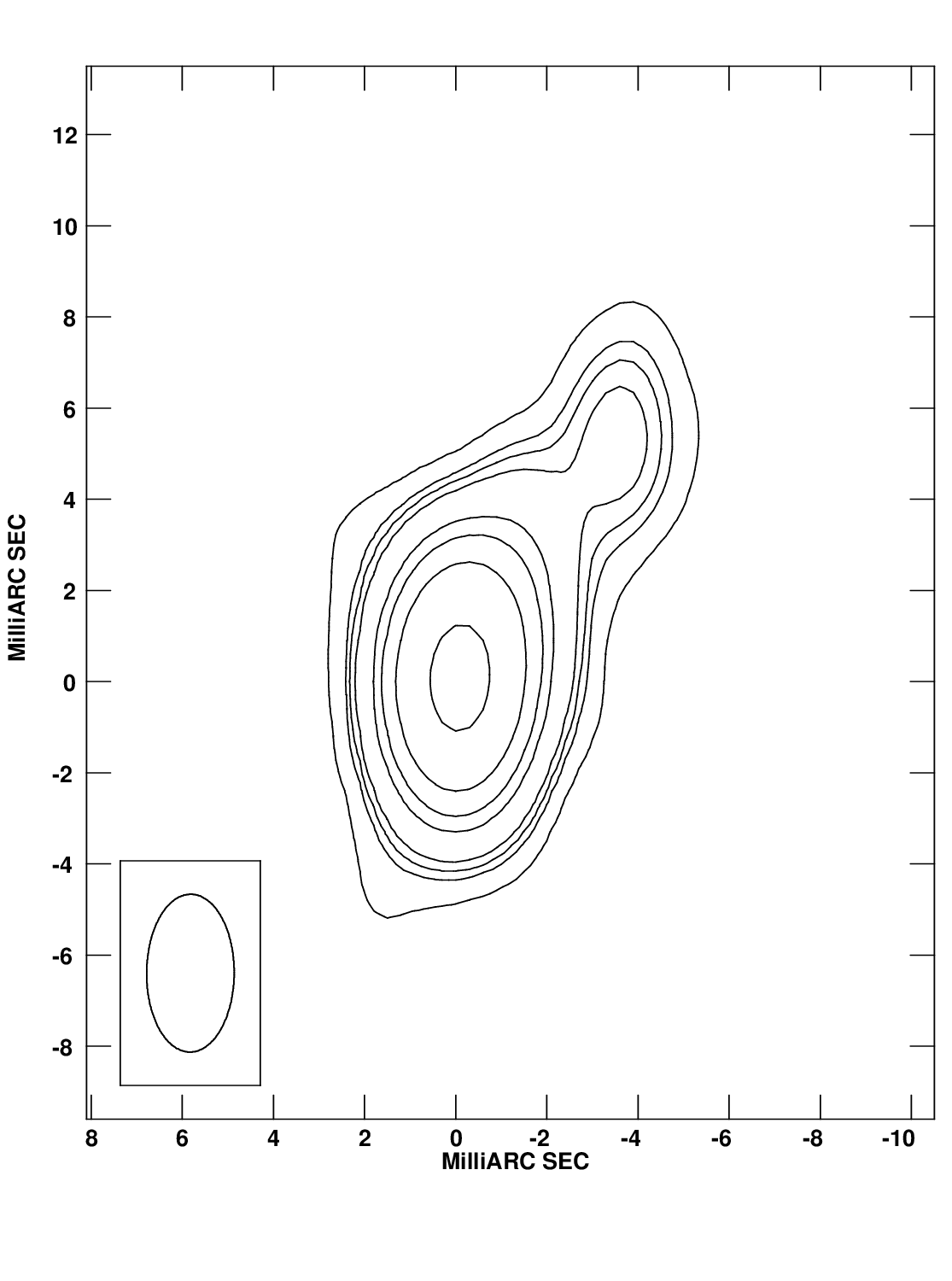}
\caption{VLBA image at 5 GHz of 1553+24. Levs are: $-$0.2 0.2 0.5 0.7 1 3 5 10 
30 mJy/beam}
\end{figure}

\begin{figure}
\epsscale{0.8}
\plottwo{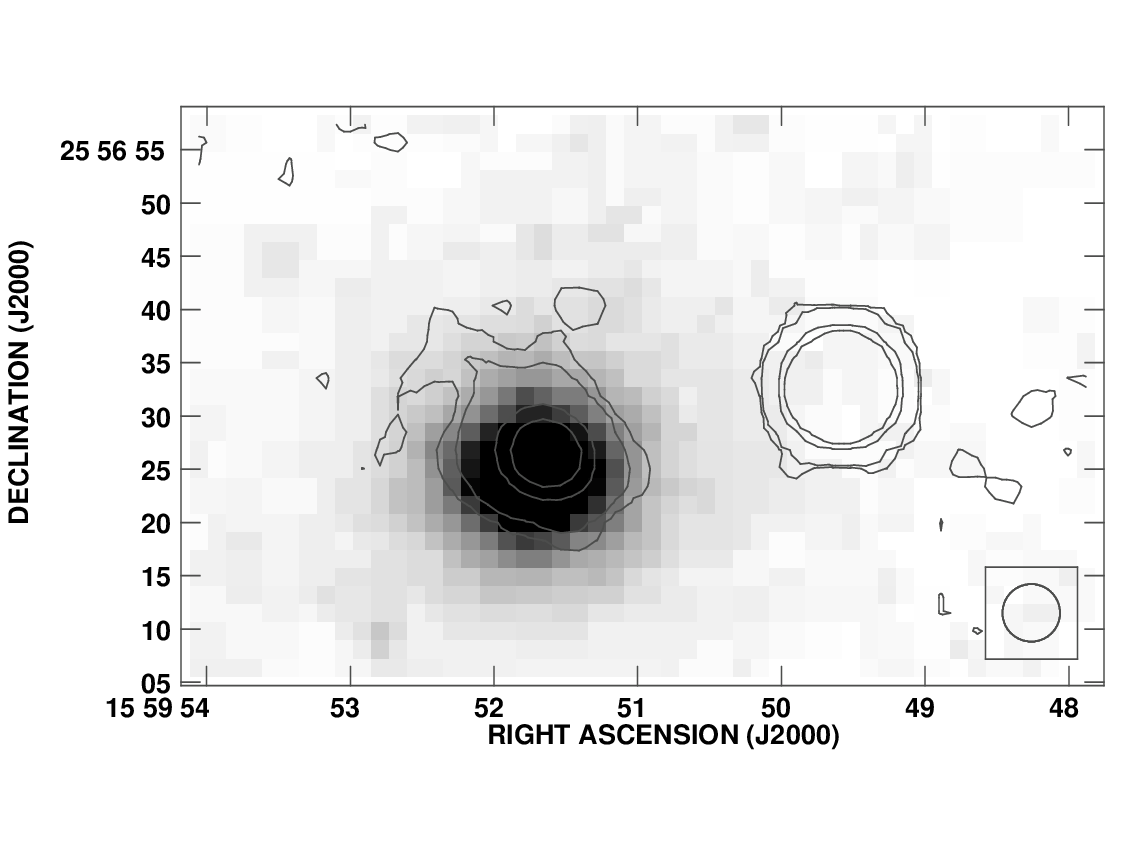}{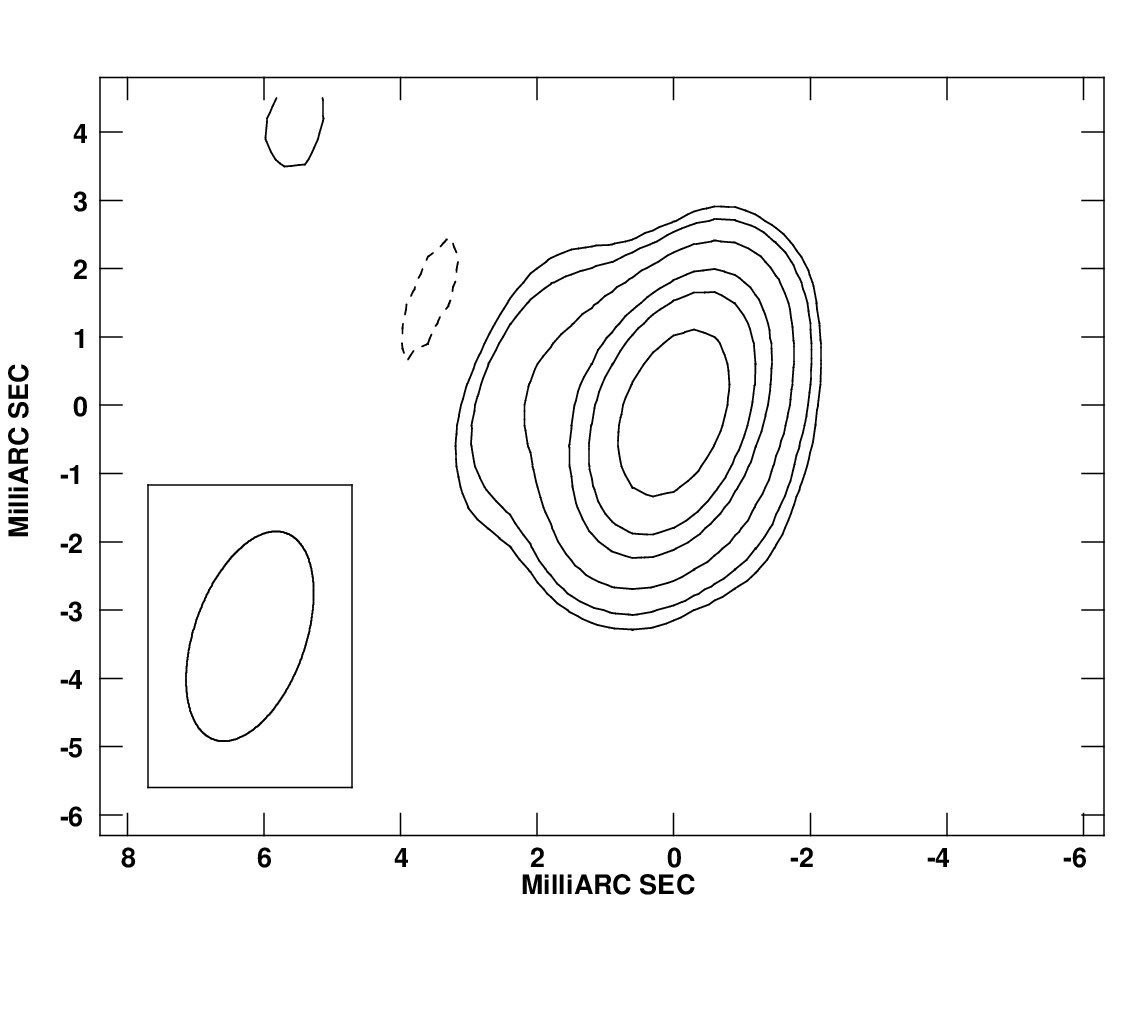}
\caption{FIRST image of 1557+26 (IC4587)
superimposed on the optical gray-scale image
(left). Levs are: 0.3 0.7 5 10 mJy/beam, and VLBA image at 5 GHz of 1557+26
(right). Levs are: $-$0.4 0.3 0.5 1 2 3 5 mJy/beam.}
\end{figure}

\clearpage

\begin{figure}
\epsscale{0.4}
\plotone{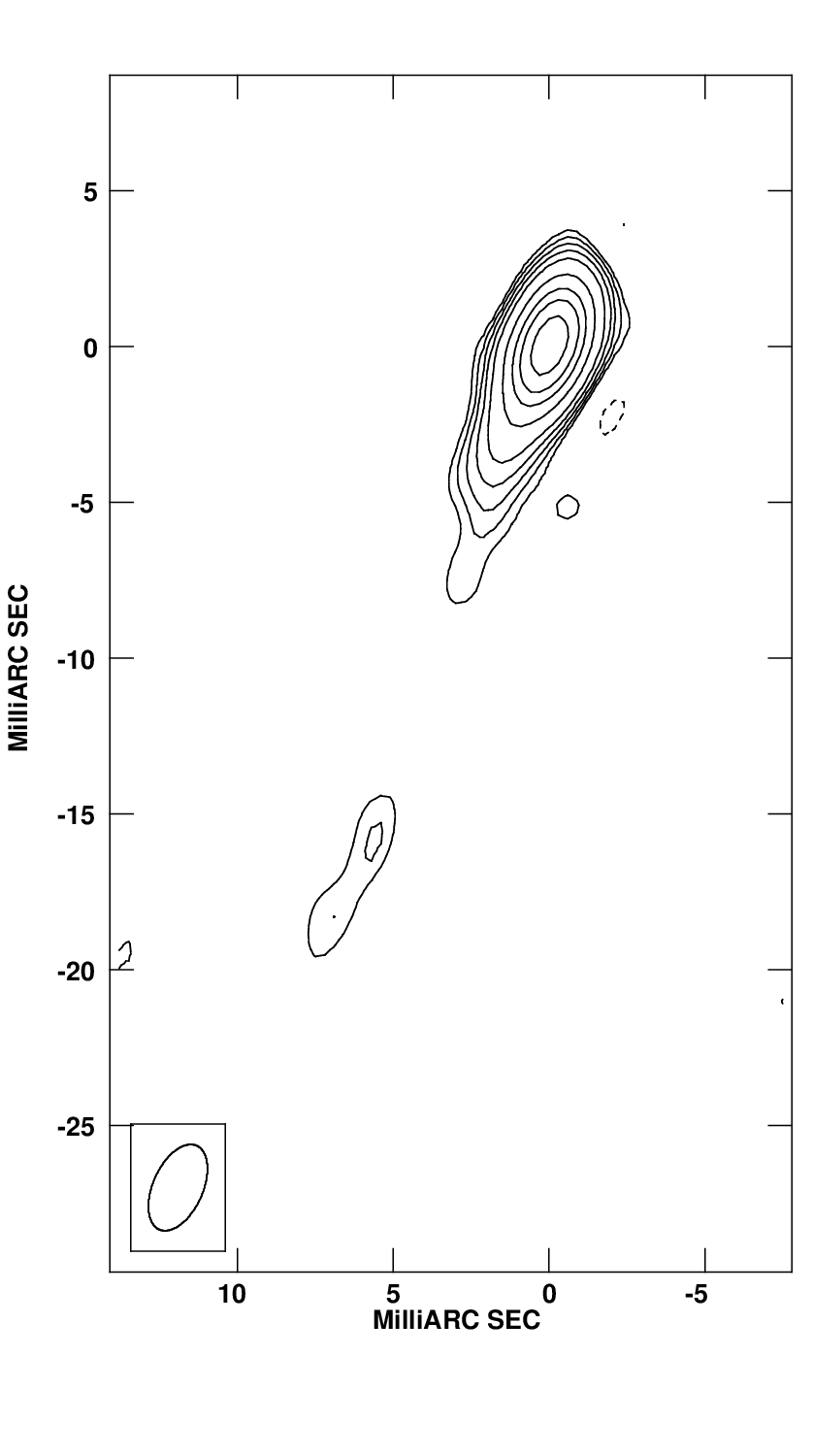}
\caption{VLBA image at 5 GHz of 1621+38 (N6137). 
Levs are: $-$0.35 0.35 0.5 0.7 1
1.5 3 5 7 10 mJy/beam.}
\end{figure}

\begin{figure}
\epsscale{0.3}
\plotone{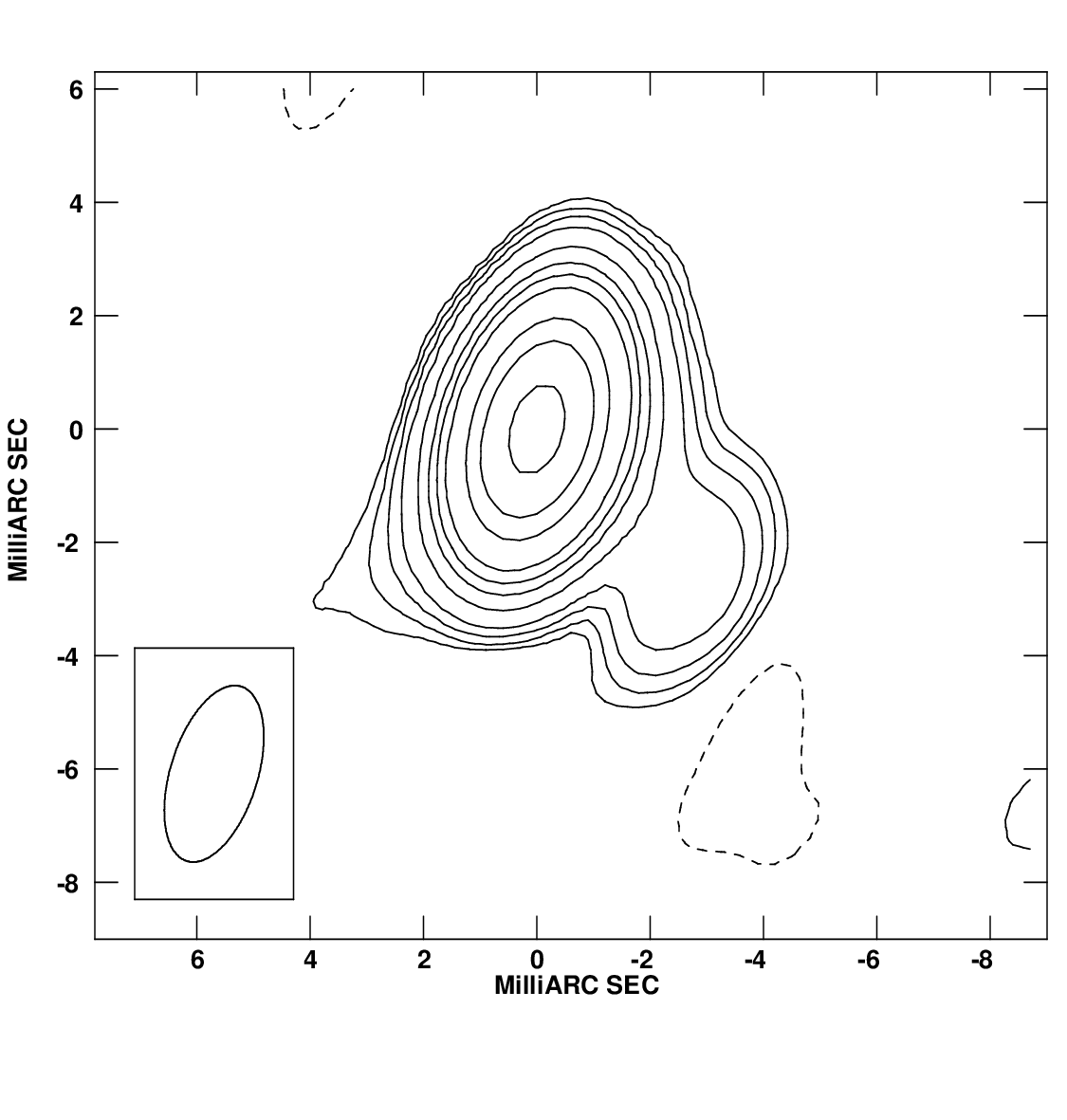}
\caption{VLBA image at 5 GHz of 1658+30 (4C30.31). 
Levs are: $-$0.5 0.5 0.7 1 1.5 3 5
7 10 20 30 50 mJy/beam.}
\end{figure}

\begin{figure}
\epsscale{0.3}
\plotone{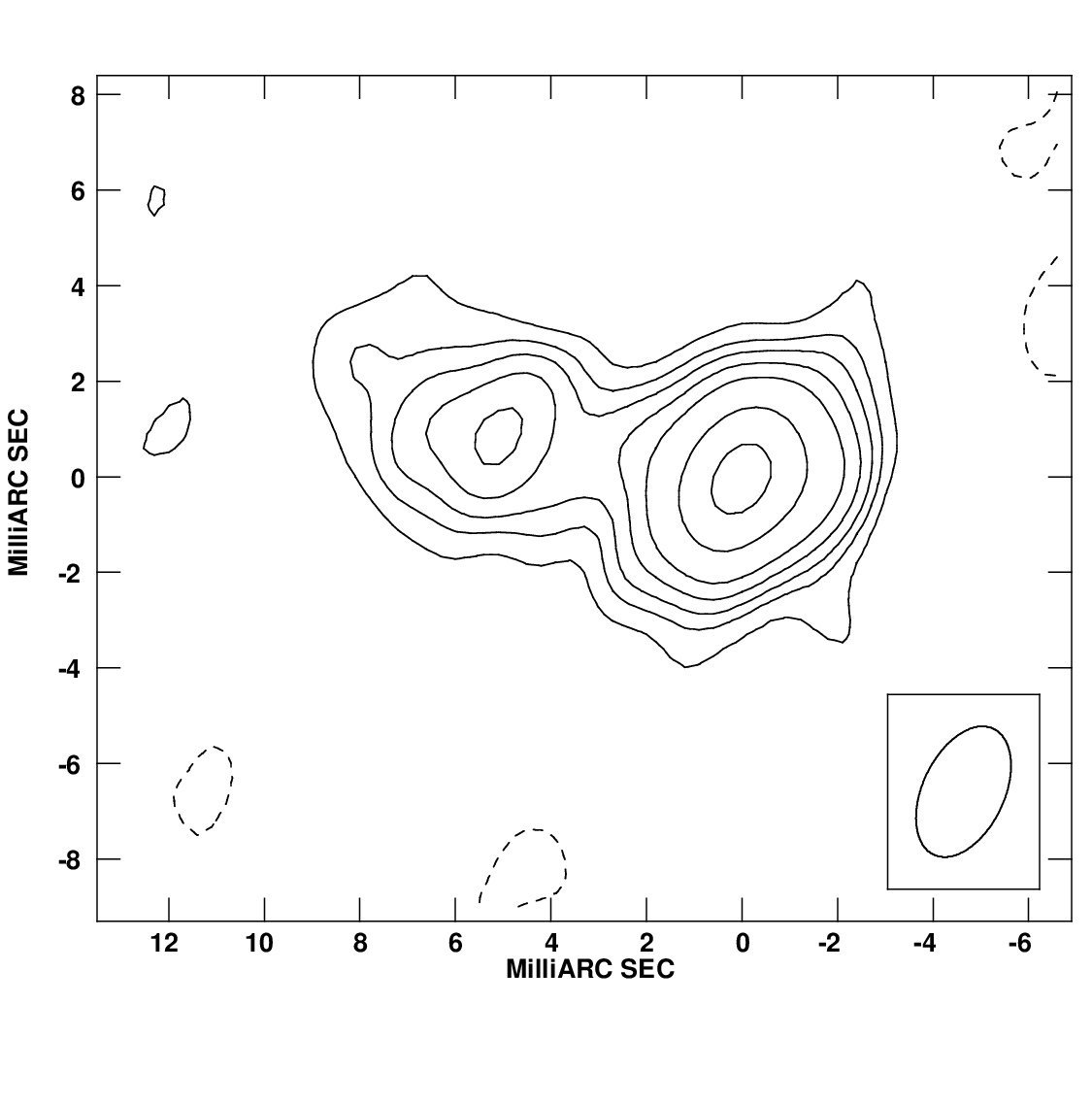}
\caption{VLBA image at 5 GHz of 1827+32A. Levs are: $-$0.3 0.3 0.5 0.7 1 1.5
3 5 mJy/beam}
\end{figure}

\begin{figure}
\epsscale{0.3}
\plotone{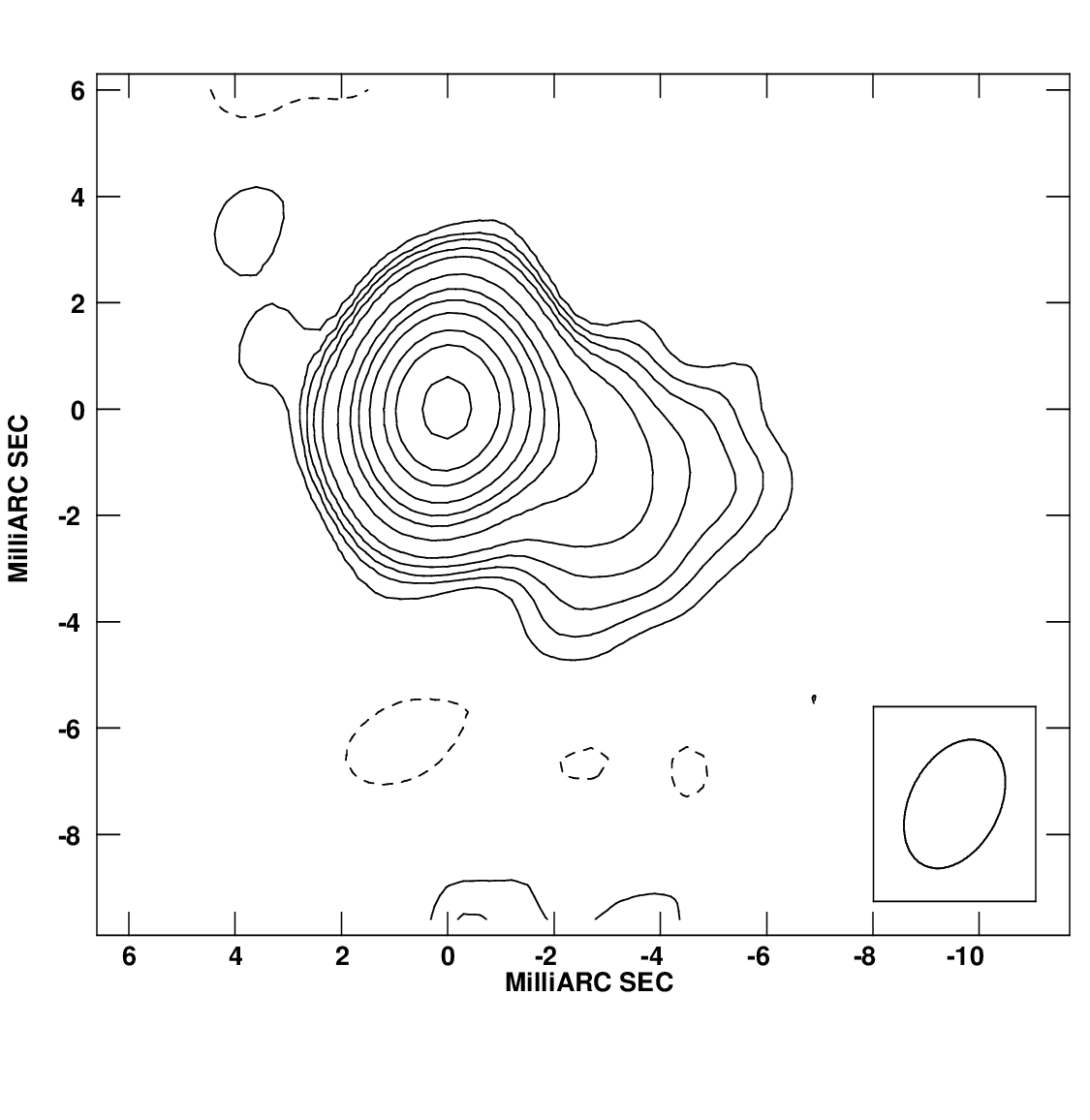}
\caption{VLBA image at 5 GHz of 1842+45 (3C 388). Levs are: $-$0.3 0.3 0.5 0.7
1 1.5 3 5 7 10 15 20 30 mJy/beam.}
\end{figure}

\begin{figure}
\epsscale{0.3}
\plotone{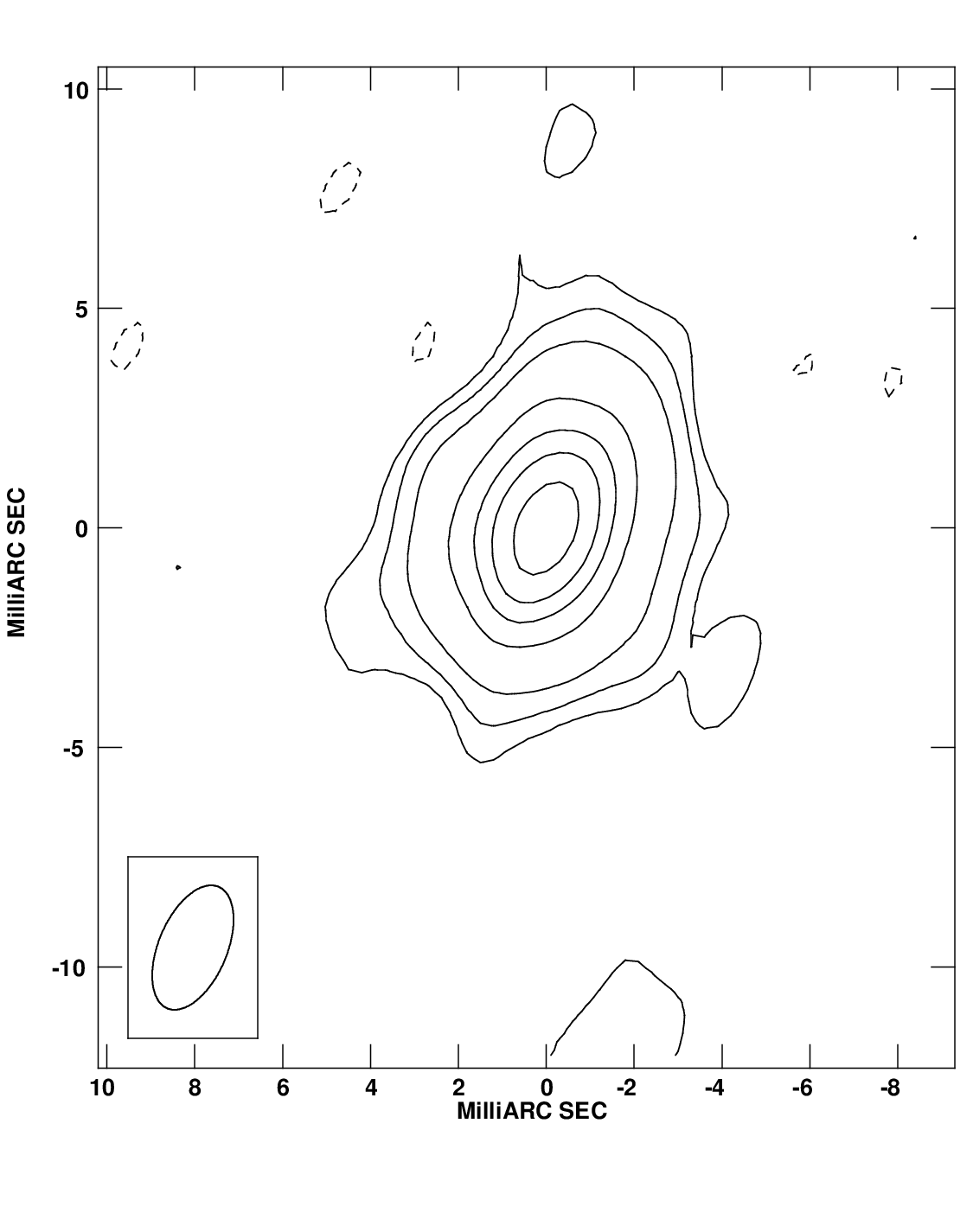}
\caption{VLBA image at 5 GHz of 2229+39 (3C 449). Levs are: $-$0.3 0.25 0.5 
1 3 5 7 10 mJy/beam.}
\end{figure}

\vfill\eject

\end{document}